\documentclass{acm_proc_article-sp}
\usepackage{soul}
\usepackage{xcolor}
\usepackage{url}
\usepackage{amsmath}
\usepackage{pstricks}    
\usepackage{pst-circ}    
\usepackage{pst-tree}    
\usepackage{pst-node}    
\usepackage{subcaption}
\usepackage{csvsimple}
\usepackage{authblk}

\newcommand{\reffig}[1]{Figure~\ref{#1}}
\newcommand{\refeq}[1]{(\ref{#1})}
\newcommand{\reftab}[1]{Table~\ref{#1}}

\newcommand{\refsec}[1]{Section~\ref{#1}}
\newcommand{\ie}{i.e.,~}
\newcommand{\eg}{e.g.,~}
\newcommand{\etal}{\textit{et al.,~}}
\begin{document}

\title{Power Side Channels in Security ICs: Hardware Countermeasures}

\author[1]{\aufnt Lu Zhang}
\author[2]{\aufnt Luis Vega}
\author[3]{\aufnt Michael Taylor}
\affil[ ]{\aufnt Computer Science and Engineering}
\affil[ ]{\aufnt University of California, San Diego}
\affil[ ]{\aufnt {\{luzh\textsuperscript{1},
                    lvgutierrez\textsuperscript{2},
                    mbtaylor\textsuperscript{3}\}@eng.ucsd.edu}}

\maketitle
\begin{abstract}
Power side-channel attacks are a very effective cryptanalysis technique that
can infer secret keys of security ICs by monitoring a chip's power consumption.
Since the emergence of practical attacks in the late 90s, they have been a
major threat to many cryptographic-equipped devices including smart cards,
encrypted FPGA designs, and mobile phones. Designers and manufacturers of
cryptographic devices have in response developed various countermeasures for
protection. Attacking methods have also evolved to counteract resistant
implementations. This paper reviews foundational power analysis attack
techniques and examines a variety of hardware design mitigations. The aim is to
highlight exposed vulnerabilities in hardware-based countermeasures for future
more secure implementations.
\end{abstract}





\section{Introduction}
Side-channel attack scenarios take advantage of information leaked from
channels other than the main communication channel and discover small but
critical secrets from the leakages. Attacks target chips that incorporate
cryptographic functions in order to reveal the secret keys using side-channel
analysis techniques. Well-known side channels include variations in
execution time, power consumption, and electromagnetic emission during
different encryption and decryption steps. Because of the non-invasive
properties of side-channel attacks they are often easy to mount without
expensive equipment, and hard to be distinguished from normal chip operations.
Kocher \etal \cite{kocher1999differential} in 1999 presented practical
approaches to performing power analysis on cryptographic devices. Two kinds of
methods were developed to extract secret keys: \textit{Differential Power
Analysis (DPA)} and \textit{Simple Power Analysis (SPA)}. The DPA method
employs statistical analysis using many power measurements, and the SPA is
applicable when the leak is so evident that simple analysis techniques such as
visual inspection can disclose secrets.

\vfill\eject

Modern cryptographic ciphers such as AES and RSA are designed to resist
adversaries that supposedly have knowledge of both plaintext and ciphertext
data. For example, to break AES, one must exhaustively enumerate all possible
cryptographic keys which is computationally prohibitive. The power analysis
techniques circumvent such difficulties by intercepting \textit{intermediate
values} that are calculated for encryption or decryption processes. The
intermediate values are much shorter in bit length than either the plaintext or
the ciphertext, allowing the adversary to take a \textit{divide-and-conquer}
strategy to recover portions of the key separately. Power analysis attacks can
be successfully mounted on a variety of cipher implementations. This paper
primarily uses the AES (Advanced Encryption Standard) block cipher as an
example for power analysis discussions due to its wide adoption. The appendix
has an introduction of the AES round functions and the cipher implementation
choices. The AES standard is specified in \cite{pub2001197}.

This paper focuses on the differential power analysis because it is much harder
to defend against than the simple power analysis attacks. The major
contribution is to evaluate emerging hardware countermeasures that have been
developed since the disclosure of power side channels and summarize their
potential vulnerabilities. The rest of this paper is organized as follows.
Section 2 briefly introduces the concept of power traces and how to set up an
attack. Section 3 reviews basic methods for differential power analysis.
Section 4 surveys the most commonly used power models. Section 5 presents
metrics for evaluating the effectiveness of attacks or defenses. Section 6
outlines various directions for mitigations. Section 7 discusses a variety of
hardware-based countermeasures, and section 8 summarizes the paper and proposes
other promising solutions.

\section{Power Traces}

Modern cryptographic algorithms are typically implemented in an integrated
circuit (IC) chip that consists of numerous logic gates composed of CMOS
(Complementary Metal-Oxide Semiconductor) transistors. Transistors
\textit{switch} on and off to represent the logic function of a gate and these
transitions draw electric current from the chip's power supply. The power
analysis attack is based on the fact that the overall chip's power variation
reflects the aggregated switching activity of each gate.

The first step for power analysis attacks is to acquire one or more
\textit{power traces} from the target device. A power trace
(\reffig{fig:powertrace}) is a digitized sampling sequence of instantaneous
power consumption values over a period when a series of cryptographic
operations take place, for example, the first round of the AES. Once a power
trace has been acquired some optional digital signal processing can be used to
improve the trace quality.

\subsection{The Measurement Setup}

Many cryptographic chips require a single constant voltage supply, \eg 5V,
3.3V or 1.8V, often denoted as $V_{dd}$. The current drawn from $V_{dd}$ by the
chip is a time-variant value $I_{dd}(t)$ in Amperes that sinks to the ground
lines of the chip. $P(t) = V_{dd} \cdot I_{dd}(t)$ describes the instantaneous
power in Watts of the whole chip.

\begin{figure}
  \centering
  \includegraphics[width=.45\textwidth]{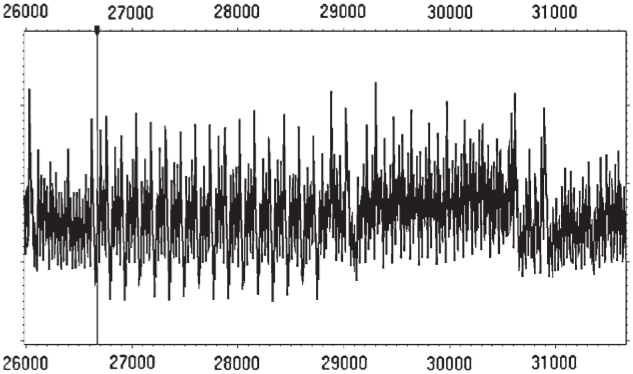}
  \caption{An Example Power Trace in \cite{kocher2011introduction}. The vertical line marks the time of the first S-Box output.}
  \label{fig:powertrace}
\end{figure}

The most commonly used sampling device is an oscilloscope that
measures voltages. A small resistor $R$ is usually inserted into the $V_{dd}$
or ground line, and the voltage drop $V_R$ across the resistor is sensed by
probes of the oscilloscope and stored. The voltage drop is proportional to the
power consumption: $V_R = I_{dd}(t) \cdot R \propto P(t)$.
\reffig{fig:powertrace} shows an example power trace representing the first
round of an AES-128 encryption on a smart card. From the time marker, it is
visible that there are 16 spikes happening in sequence, indicating the
microcontroller is working on state bytes one after another. Recorded power
traces are represented by the matrix in \refeq{eq:tracematrix}.

\begin{equation} \label{eq:tracematrix}
\mathbf{P} =
  \begin{pmatrix}
  P_1(1) & P_1(2) & \cdots & P_1(T) \\
  P_2(1) & P_2(2) & \cdots & P_2(T) \\
  \vdots & \vdots & \ddots & \vdots \\
  P_N(1) & P_N(2) & \cdots & P_N(T) \\
  \end{pmatrix}
\end{equation}

In $\mathbf{P}$ each row represents one power trace measured
within a period $T$, for the same sequence of operations performed on $N$
different sets of data. For example, the first round of AES-128 encryptions
using the \textit{same} cipher key. The instantaneous value $P_n(t)$ in the matrix is
digitized, and stored as binary integers. Because of sampling, the time values
represented in the matrix are discrete. One important property that must be
preserved between rows of the matrix is the \textit{time consistency}, meaning
the values of each column must preserve the same time offset relative to the
start of sampling.

In practice, many measurement setups have to be carefully performed to
get clear power traces. For example, a highly stable supply voltage is
preferred. The serial resistor should be inserted close to the target chip's
supply or ground pins, not the circuit board or device level ones. Modern SoCs
usually have several power supplies for different parts of its internal
structures, which, on the other hand, all share the same ground. Some chip I/O
pins being pulled down are also electrically grounded and may shunt some amount
of return currents. In such cases, the ground lines are noisier than the power
line that mainly supplies the internal cipher module. Unless specially noted
this paper assumes attacks are executed in the power line, and higher values of
matrix $\mathbf{P}$ indicate higher power consumptions from the source $V_{dd}$.


\subsection{The Signal Processing} \label{sec:signal-processing}

To suppress noise and improve time consistency, optional signal processing
can be performed on the raw power traces. Noise signals can be filtered by
applying appropriate filters. Time consistency is often preserved by a
\textit{trigger} signal \cite{kocher2011introduction} that indicates the
beginning of certain functions, for example, the first AES round. This signal
can trigger the oscilloscope to start sampling, and the resulting
power traces will be well aligned. In case such a trigger signal is not available,
\textit{trace alignment} techniques are employed. A simple correlation test
could be helpful to find the time shift $\tau$ that minimizes the differences
between $P_i(t)$ and $P_j(t + \tau)$. Occasionally more complex alignment
methods are needed in the presence of hardware countermeasures
\cite{clavier2000differential} or for a very noisy device such as a smartphone
\cite{nakano2014pre}. The power trace matrix shown in \refeq{eq:tracematrix}
can be considered as filtered and aligned, and each row corresponds to the
encryption process for one data block of 16 bytes.


\section{Differential Power Analysis}


The first differential power analysis introduced by \cite{kocher1999differential}
uses a method called \textit{Difference of Means (DoM)}. Later versions using
more effective statistical tools have been developed to improve the performance
of differential power analysis. One easy to mount and very powerful attack
calculates correlation coefficients and thus it is often called the
\textit{correlation power analysis (CPA)}. This section mainly introduces these
two methods because hardware countermeasures often use them for the
evaluations. A few advanced attacking techniques are also briefly introduced.

\subsection{Difference of Means}


The differential power analysis published by \cite{kocher1999differential}
targeted a DES (Data Encryption Standard) cipher, and later the same authors
extended their work to AES \cite{kocher2011introduction}. They introduced a
\textit{known-plaintext} attack: the adversary has a set of plaintext data,
encrypted by an AES cipher using the \textit{same} secret cipher key. For each
plaintext input, the adversary can observe a power trace and for all the
$N$ inputs (e.g.  4,000 in \cite{kocher2011introduction}) the power traces are
stored in a matrix like $\mathbf{P}$. Each power trace represents the
encryption of a data block of 16 bytes (\reffig{fig:powertrace}). In this
scenario, the ciphertext can be used at the end to verify the correctness of the
key recovered, but for the extraction process the ciphertext is not used.

\begin{figure}
\centering
  \includegraphics[width=.45\textwidth, height=.25\textwidth]{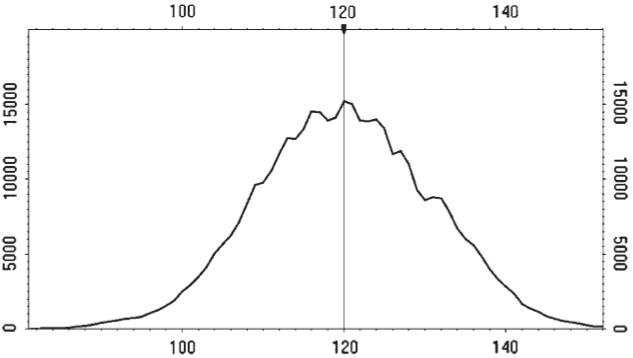}
  \caption{The Power Consumption Distribution at the Time Marker of \reffig{fig:powertrace}}
  \label{fig:gaussian1}
\end{figure}
\begin{figure}
\centering
  \includegraphics[width=.45\textwidth, height=.25\textwidth]{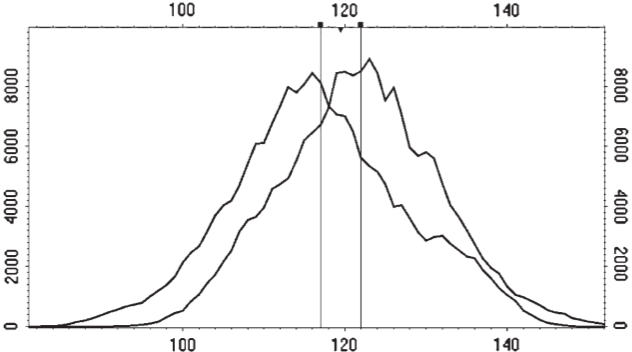}
  \caption{Distributions of the Two Subsets at the Time Marker of \reffig{fig:powertrace}}
  \label{fig:gaussian2}
\end{figure}

Observations show that the power trace values sampled at a specific point in time
(a column of the power trace matrix) is close to a \textit{normal distribution}.
\reffig{fig:gaussian1} shows the distribution at the time of
the first S-Box output in the initial AES round. The distribution is a combined
effect caused by all switching transistors of the chip and electronic noises.
Now \textit{suppose} the cipher key is known then the statistical analyze is as
follows. For each data input, the result after the first AddRoundKey and
SubBytes can be calculated. Then according to one arbitrary bit of the output,
for example, the least significant bit (LSB), separate the power traces into two
subsets. One subset has the traces where the LSB is \texttt{0}, and the other
has traces where the LSB equals \texttt{1}. Now the distributions of the two
subsets are still close to normal but with distinguishable means, shown in
\reffig{fig:gaussian2}.

The way to separate traces is known as the \textit{selection function}
\refeq{eq:sf1} for DoM based attacks. The reason for the difference of means
is circuit design details make it possible for either LSB value to consume more
power when data is being processed. Switching to the attack scenario, the
adversary, without knowing the cipher key, can do exhaustive testing of all key
hypotheses, and each time use the \textit{hypothetical} LSB to separate power
traces into two subsets. Then the adversary calculates the difference of means
as in equation \refeq{eq:dom}. When the key guess is correct, the difference of
the subsets' means usually stands out as shown in \reffig{fig:gaussian2}.  Even
if the difference is very small, it will become \textit{statistically
significant} given a sufficient number of power traces (large $N$ values in
matrix $\mathbf{P}$). If the key guess is not correct, each subset is
essentially a random sampling of the full distribution and their DoM often
vanishes as the number of trace increases.

\begin{equation} \label{eq:sf1}
  F(d_n, ~ x_k) = \mathit{LSB}( ~ S(d_n \oplus x_k) ~ )
\end{equation}

In equation \refeq{eq:sf1} and \refeq{eq:dom}, $S(\cdot)$ indicates the
SubBytes function, $d_n$ represents one data \textit{byte} of the initial
state of the $n^\mathit{th}$ power trace, and $x_k$ ($k \in [1 \cdots 256]$) is
a guessed round key \textit{byte} corresponding to $d_n$. The
$\Delta\bar{P}(t)$ is known as the \textit{differential trace}, and the
predicted byte $x_{\hat{k}}$ of the first round key is chosen by the greatest
difference of means at certain time $\hat{t}$. Thus, a successful attack also
reveals the time $\hat{t}$ when the key byte $x_{\hat{k}}$ is used. The same
DPA process is repeated for each key byte independently. Because this attack
scenario targets the S-Box output of the \textit{first} round, the recovered
round key is also the predicted cipher key.

\begin{equation} \label{eq:dom}
\begin{split}
  \Delta\bar{P}(t) = & \frac{ \sum_{n=1}^{N} F(d_n, ~ x_k) \cdot P_n(t) }{ \sum_{n=1}^{N} F(d_n, ~ x_k) } \\
                     & - \frac{ \sum_{n=1}^{N} (1 - F(d_n, ~ x_k)) \cdot P_n(t) }{ \sum_{n=1}^{N} (1 - F(d_n, ~ x_k)) }
\end{split}
\end{equation}

\begin{equation} \label{eq:domkey}
  (\hat{k}, \hat{t}) = \underset{(k, t)}{\arg\!\max} ~ \lvert \Delta\bar{P}(t) \rvert
\end{equation}


Mounted DPA attacks, in practice, are often more complex. The target bit does not
have to be the LSB, and different bits could have varying degrees of
leakages. There could also be more than one guess that leads to equally strong
spikes in the differential trace, and a decision based on equation
\refeq{eq:domkey} is hard to make. The false spikes caused by incorrect key
bytes are termed as \textit{harmonics} or \textit{ghost peaks}, which are
essentially caused by several key hypotheses all correlate to the chosen bit.
Nevertheless, such a simple method worked devastatingly well on many smart cards
when the paper \cite{kocher1999differential} was published. Later methods have
evolved to make great improvements in this classical DPA technique.

\subsection{Correlation Coefficient}
The correlation power analysis \cite{brier2004correlation} introduced
by Brier \etal was the first to explicitly use the \textit{correlation
coefficient} (also known as the \textit{Pearson Product-Moment Correlation
Coefficient}) to make decisions among key hypotheses. To deploy a correlation
analysis, the adversary needs to build a hypothetical power consumption matrix
$\mathbf{H}$, for every key guess in $\{x_1, x_2, \cdots, x_K\}$ ($K = 256$ in
case only one key byte is targeted in one attacking attempt). In $\mathbf{H}$,
each column is calculated using one value of $x_k$, and the entry $H_n(k)$ is a
modeled power value estimated by using a plaintext byte $d_n$ and the
hypothetical key byte $x_k$.

\begin{equation} \label{eq:hypomatrix}
\mathbf{H} =
  \begin{pmatrix}
  H_1(1) & H_1(2) & \cdots & H_1(K) \\
  H_2(1) & H_2(2) & \cdots & H_2(K) \\
  \vdots & \vdots & \ddots & \vdots \\
  H_N(1) & H_N(2) & \cdots & H_N(K) \\
  \end{pmatrix}
\end{equation}

\begin{equation} \label{eq:cor}
\begin{split}
  R_k(t)     & = \frac{ \sum_{n=1}^{N} (H_n(k) - \bar{H}(k)) \cdot (P_n(t) - \bar{P}(t)) }{ \sqrt{ \sum_{n=1}^{N} (H_n(k) - \bar{H}(k))^2 \cdot \sum_{n=1}^{N} (P_n(t) - \bar{P}(t))^2} } \\
             & with: \bar{H}(k) = \frac{\sum_{n=1}^{N} H_n(k)}{N}, ~ \bar{P}(t) = \frac{\sum_{n=1}^{N} P_n(t)}{N}
\end{split}
\end{equation}

Therefore, the computing of $H_n(k)$ depends on how the adversary models the
power consumptions according to the cipher's intermediate values. The paper
\cite{brier2004correlation} suggested a \textit{Hamming Distance} model, which
is the count of different bits between two bit-vectors. The model could target
the state register of the AES cipher, and assumes its power consumption has a
linear relationship with the number of flipped bits between two successive
states. For the known-plaintext attack scenario, the two successive state
values can be the results of AddRoundKey and SubBytes in the first round. Hence
$H_n(k)$ is computed by the Hamming distance between $(d_n \oplus x_k)$ and
$S(d_n \oplus x_k)$.

\begin{equation} \label{eq:corkey}
  (\hat{k}, \hat{t}) = \underset{(k, t)}{\arg\!\max} ~ \lvert R_k(t) \rvert
\end{equation}


The correlation coefficient $R_k(t)$ ($[-1.0, 1.0]$) between each column
$\mathbf{H}(k)$ of $\mathbf{H}$ and each column $\mathbf{P}(t)$ of $\mathbf{P}$
is computed as in \refeq{eq:cor}. The greatest value of $\lvert
R_{\hat{k}}(\hat{t}) \rvert$ given by \refeq{eq:corkey} predicts the secret key
byte $x_{\hat{k}}$ and the time $\hat{t}$ of its leakage.




The explained DPA and CPA approaches are suitable for a known-plaintext
scenario. For encryption power traces with \textit{ciphertext-only}, the
attacking process needs to be adjusted. For example, the target intermediate
value becomes the \textit{last round S-Box input}, and the attacker tries to
predict the \textit{last round key}. Since the key expansion for AES-128 is
invertible with any round key, having the last round key also reveals the
cipher key. Similar attack flows can be derived for the decryption process as
well. The first or the last round of the block cipher is often targeted by DPA
because the data complication is sufficiently weak for some of the intermediate
values to be easily enumerated. It is possible that an attack does not uniquely
predict a key, or the most likely choice is not the right one. In such cases, a
list of candidates can be supplied to some key enumeration and validation
steps. Nevertheless, the attack results should have reduced the remaining
brute-force effort to feasible levels.


\subsection{Advanced Power Analysis Attacks}
Since the debut of practical power analysis attacks, many evolved approaches
have been proposed to improve either DPA or SPA. For instance, the selection
function for the DoM method can assign weights to different traces or divide
traces into more than two categories. Using functions other than the ordinary
averaging could be useful when data sets have unusual statistical
distributions. Brief introductions about a few other major improvements to
power analysis techniques are as follows.

\textbf{Higher-Order DPA:} The attacks introduced in the previous section only
analyze the statistics of samples at \textit{one} point in time, \ie correlate
$\mathbf{H}(k)$ with each column of $\mathbf{P}$ independently. This is known
as the \textit{first-order} attack. \cite{kocher1999differential} also
introduced \textit{higher-order} DPA that works on samples at multiple
times within $\mathbf{P}$ and their cross-correlations. The \textit{attacking
order} is roughly the number of intermediate values of different time being
considered in one attack. Practical results (\cite{messerges2000using} and
\cite{oswald2006practical}) have reported that higher-order attacks can
compromise implementations with certain countermeasures.

\textbf{Collision Attacks:} A collision attack \cite{schramm2004collision}
requires a pair of encryption runs with two known and different inputs. It
checks if the two encryption runs compute some equal intermediate
values, known as \textit{intermediate value collisions}. The essential
idea is that for two different inputs, a collision cannot occur for all
key values, but only for a limited subset of keys. Hence, the collisions
reduce the guessing space and possibly uniquely identify the cipher
key. Adversaries using this attacking method may prefer to choose some
plaintext inputs for efficient collision detection, and such a scenario is
called the \textit{chosen-plaintext attack}. In practice, one pair of chosen
plaintexts might only allow the recovery of one key byte. A few more
measurements are required to predict the whole cipher key, but the number of
required power traces is usually significantly reduced. Enhanced collision
attack using correlation analysis has also been developed by
\cite{moradi2010correlation}.

\textbf{Profiled Attacks:}
This series of attacking methods aim to solve the problem when only one or two
power traces of the victim device are observable. The template attack
\cite{chari2003template} builds an accurate power consumption model using an identical or
similar device as the victim. The attack consists of two steps. First a
\textit{profiling} phase is performed by precisely characterizing power traces
as a \textit{multivariate normal distribution}. Next during the on-site attack
a \textit{key extraction} phase is executed using the available templates and
key hypothesis testing on a \textit{single} power trace observed from a victim
device. The researchers successfully used the template attack to break an RC4
stream cipher and a DES block cipher. Schindler \etal
\cite{schindler2005stochastic} introduced the stochastic models for the device
profiling. It achieves the efficiency of the template attack in the key
extraction phase but requires far fewer measurements in the profiling phase.




\textbf{Algebraic Side-Channel Attacks:} The algebraic side-channel attack
\cite{renauld2010algebraic} combines template attacks and algebraic
cryptanalysis. Besides building templates, it also exploits the information
leakage of many cipher rounds (not limited to the first or last round) to build
a boolean satisfiability problem to solve. The authors managed to break a
PRESENT block cipher with the observation of a single power trace. A later
improved version \cite{mohamed2012improved} also successfully attacked an
AES-128 implementation.


\section{Power Models} \label{sec:power-models}


%
%


\begin{figure}
\centering
  \includegraphics[width=.45\textwidth]{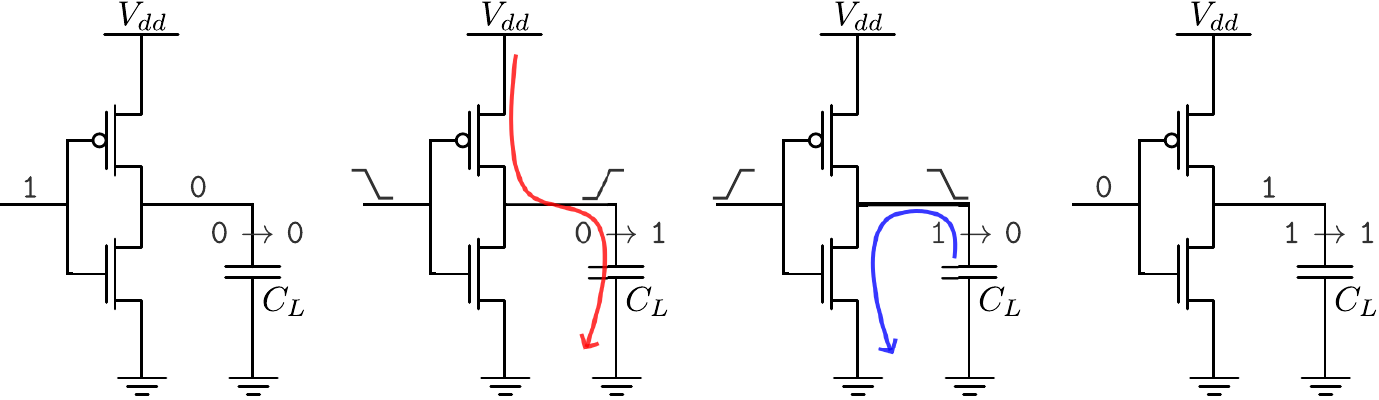}
  \caption{CMOS Inverter Transitions}
  \label{fig:transitions}
\end{figure}

Most power side-channel attacks exploit the \textit{linear relations} between
the observed power traces and the hypothetical power consumption that is
dependent on intermediate values the circuitry processes.  Analysis of
transistor-based logic gates reveals a clear dependency between its output
value and the power it consumes. \reffig{fig:transitions} illustrates a basic
CMOS inverter gate and simplified models of its transition activities.  Its
power consumption includes three parts: the \textit{static power}\footnote{The
static power is also called the \textit{sub-threshold leakage power}. To avoid
confusion with the information leakage in this context, this paper refers to it
as the static or just sub-threshold power.}, $P_\mathit{static}$, the
\textit{short-circuit power}, $P_\mathit{short}$ and the \textit{switching
power}, $P_\mathit{switch}$. The sum of the latter two components is also known
as the \textit{dynamic power} $P_{dyn}$. According to \cite{rankl2010smart},
the semiconductor technology used for the fabrication of smart card
microcontrollers lies in the range of 0.18-$\mu$m to 90-nm. And the dominant
factor of power consumption for a technology node in this range is the dynamic
power \cite{moradi2009vulnerability}. This paper only considers the dynamic
power, since most power side-channel attacks rely on this portion.  Attack
results based on sub-threshold power are still preliminary
\cite{moradi2014side}.

The capacitor $C_L$ in \reffig{fig:transitions} is a lumped model including the
transistor's intrinsic capacitance, its load capacitance, and the parasitic
capacitance of the interconnections. When its output switches from \texttt{0}
to \texttt{1}, $C_L$ gets charged by a current from
$V_{dd}$. The switching power the gate consumes is proportional to
$\alpha \cdot f \cdot C_L \cdot V_{dd}^2$, where $\alpha$ is the active factor
of \texttt{0} $\to$ \texttt{1} transitions in one clock cycle, and $f$ is the
clock frequency of the circuit. When the output switches from \texttt{1} to
\texttt{0}, $C_L$ discharges and ideally there are no current flows from the
power supply. However in both \texttt{0} $\to$ \texttt{1} and \texttt{1} $\to$
\texttt{0} transitions there is a momentary short-circuit current flowing from
$V_{dd}$ to the ground when both the transistors are conducting. The \texttt{0}
$\to$ \texttt{0} and \texttt{1} $\to$ \texttt{1} transitions do not incur power
variations if not considering the sub-threshold current. Accordingly, if
measure the power consumption in the $V_{dd}$ line, \texttt{0} $\to$ \texttt{1}
transitions contribute the most, and then the \texttt{1} $\to$ \texttt{0}
cases. Now consider an \texttt{AND} which computes $y = a \cdot b$ and its
switching activity shown in \reftab{tab:switching} (just the $a$, $b$, and
\texttt{AND} columns). Given a uniform distribution of switching probabilities
for $a$ and $b$, obviously the probabilities for the four output switching activities
are not uniform anymore. Assume $a$ is a fixed secret bit (like a key) and $b$
is a random bit (like a plaintext or ciphertext). If $a$ is \texttt{0} (\ie the
four \texttt{0} $\to$ \texttt{0} rows of $a$), $y$ has neither \texttt{0} $\to$
\texttt{1} nor \texttt{1} $\to$ \texttt{0} switching; otherwise if $a$ is
\texttt{1}, $y$ has one \texttt{0} $\to$ \texttt{1} and one \texttt{1} $\to$
\texttt{0} switching (\ie in the four \texttt{1} $\to$ \texttt{1} rows of $a$).
Therefore, by knowing $b$ and monitoring the mean power consumption of the
\texttt{AND} gate it is possible to deduce what value $a$ has. In contrast, the
\texttt{XOR} function still has uniform output switching activities and does
not easily reveal the value of $a$ in this way.

Scaling up this simple example, a whole cryptographic circuit's power
consumption is the aggregation of power consumed by all its logic gates. For
AES, the AddRoundKey function can be implemented only using \texttt{XOR} gates,
but the finite field inversion of the S-Box must employ logic functions like
\texttt{AND} or \texttt{OR} (\cite{wolkerstorfer2002asic}). Hence power
analysis attacks work and often favor the S-Box because of the distortions in
the odds of switching activities make distinguishable power profiles between
the right and wrong key guesses. With a chosen power analysis method, \eg the
CPA, the quality of mapping intermediate values to power consumptions has an
important impact on its effectiveness. This section gives a brief survey of
commonly used mapping methods, also known as the \textit{power models}.



\textbf{Single Bit:} Power modeling using a single bit is the most basic
approach based on the principle explained using the \texttt{AND} gate.
Nevertheless, its effectiveness had been proved by the classical DoM based DPA
attacks. Different target bits have been observed showing a different amount of
leakages. The differences root from variations in data paths that signals
travel along, and the ghost peak phenomena happen because multiple keys can
correlate to the same bit.


\textbf{Hamming Weight:} The Hamming weight power model is a multi-bit
extension of the single bit model. The weight is simply the number of
\texttt{1}s in a bit vector. For this model to work effectively, it should
assume all the bits are in a constant state (either all \texttt{0}s or all
\texttt{1}s) before a target value appears. Therefore the Hamming weight
of an intermediate value is roughly linear to the power consumed to process it.
Messerges \etal \cite{messerges2000using} first introduced this model to power
analysis attacks and found the fact that many smart card processors around that
time (year 2000) precisely exhibited these characteristics. The reason is
many microcontrollers precharge their buses before carry the real data, and
this even leads to simple \texttt{MOV} instructions leak the Hamming weights of
their operands \cite{mayer2000smartly}. Leakage of Hamming weight is especially
severe for look-up table based S-Box implementations because the I/O ports of
memories usually have much higher capacitance than other cells.

\textbf{Hamming Distance:} The Hamming distance is the count of
different bits in two bit-vectors, and the model was first used by the correlation
power analysis in \cite{brier2004correlation}. It does not require a priori
state of all \texttt{0}s or all \texttt{1}s before the target value appears.
But it supposes the adversary to guess two successive values of a register
to calculate the Hamming distance between them. The CPA attack
using the AddRoundKey and the SubBytes outputs exemplifies its use. It assumes
\texttt{0} $\to$ \texttt{1} and \texttt{1} $\to$ \texttt{0} transitions equally
contribute to the power consumption. Though still not very accurate, its
applicability has been proved by many practical attacks and widely used
as a generic power model.

\textbf{Switching Distance:}
Peeters \etal \cite{peeters2007power} introduced a power model called the
\textit{switching distance}. It exploits the fact that \texttt{0} $\to$ \texttt{1}
and \texttt{1} $\to$ \texttt{0} transitions draw a different amount of
current from the supply. For an inverter, a \texttt{0} $\to$ \texttt{1}
switching draws some current from the $V_{dd}$ for charging the $C_L$ and
the short-circuit effect while during a \texttt{1} $\to$ \texttt{0} switching
only some short-circuit current appears. Hence, the switching distance is
defined as a tunable factor $\delta$, indicating the power consumption
difference between \texttt{0} $\to$ \texttt{1} and \texttt{1} $\to$ \texttt{0}
transitions.

\textbf{Toggle Count:}
The above models are good approximations for registers that usually change
states only once in a clock cycle. Mangard \etal \cite{mangard2005successfully}
noted the fact that glitches in combinational circuits are the dominant factor
of power consumptions in a clock cycle. Glitches are unwanted output toggles
before it settles due to different arrival times of input signals and the
gate's function. Therefore, this model evaluates the number of gate toggles in
the S-Box circuit instead of its final output. It works the best on
hardware-arithmetic based S-Box implementations due to the complex
combinational data paths. The model supposes the attacker knows certain design
details of the cipher, for example, the gate-level circuit netlist, and builds an
accurate power model using circuit simulations.

\textbf{Power Simulation:}
More accurate modeling of the CMOS gates is obtained by advanced power
simulation. Commercial IC design tools provide power simulation capabilities
at different levels. From high to low, simulations can be categorized into
behavioral, netlist, layout and analog levels. Accordingly in the same order
the simulation requires greater details of the IC design and hence provides
more accurate results.

\textbf{Profiled Model:} The profiling of template and stochastic attacks can
be viewed as power modeling as well. Different from hypothetical modeling, the
profiling utilizes real measurements performed an identical or similar
device as the victim. While other attack forms consider noises as hindrance,
profiled attacks precisely model noises as well. The probability distributions
of noises using different cryptographic keys are characterized during the
profiling. Power side-channel analysis usually considers profiled attacks as
the strongest form.


The power consumption values predicted by hypothetical models do not have to be
in Watts as long as they preserve a (partial) linear relationship to the actual
data-dependent power. It is reasonable that the better a model fits the
actual power consumption of the chip, the more effective an attack becomes, and
obviously better power modeling requires more knowledge of the adversary about
the victim device. A comparative study in \cite{mestiri2013comparative} showed
that a Hamming weight model failed while the Hamming distance and switching
distance models succeeded in the same CPA attack setup. Advanced power models
such as toggle counts and power simulations are not infeasible since present
time witnesses a growth in open-source hardware and hardware as IP businesses.
It would be a difficult problem to distinguish benign and malicious users.



\section{Metrics} \label{sec:metrics}

Fair evaluation methods for side-channel attacks and countermeasures are useful
to compare their quality. The most commonly used metric for attacks and
countermeasures is the \textbf{number of power traces}, \ie the value $N$ of
$\mathbf{P}$. However this metric is subject to the power monitoring method,
the quality of tools and the effectiveness of attacks. Mangard \etal
\cite{mangard2008power} developed a more analytical way to evaluate the
vulnerability of a device to power analysis attacks. They model the power
consumption as the sum of the \textit{exploitable power} $P_{exp}$, the
\textit{switching noise} $P_{sw.noise}$, the \textit{electronic noise}
$P_{el.noise}$, and the constant component $P_{const}$. The four components are
independent of each other, and their effects are additive.

\begin{equation} \label{eq:ptotal}
  P_{total} = P_{exp} + P_{sw.noise} + P_{el.noise} + P_{const}
\end{equation}

The $P_{exp}$ is the component that depends on intermediate values (\eg an
S-Box output during an encryption) that the attacker is looking for, or, in
other words, the power that relates to secrets. The exploitable power is
intrinsic to a cipher implementation and independent of attacking methods. For
attackers, they try to use power models
to approximate $P_{exp}$, and from the defensive perspective this $P_{exp}$ should
be carefully evaluated and minimized. The switching noise $P_{sw.noise}$
results from transistor switches that are independent of the sensitive values,
for example, another processor core doing irrelevant tasks. The electronic
noise $P_{sw.noise}$ describes variations caused either internal or external
sources. The $P_{const}$ is the constant part such as the static power of the
cryptographic IC. Given an attack scenario, the signal-to-noise ratio is
defined by \refeq{eq:snr}, and the $Var(\cdot)$ function calculates the
variances of its operand.


\begin{equation} \label{eq:snr}
  \mathit{SNR} = \frac{Var(\mathit{Signal})}{Var(\mathit{Noise})} = \frac{Var(P_{exp})}{Var(P_{sw.noise} + P_{el.noise})}
\end{equation}

The researchers also derive the correlation coefficient $R_k(t)$ in terms of
$P_{exp}$ and the $\mathit{SNR}$ as equation \refeq{eq:corsnr}. The $\rho(X,
Y)$ function calculates the correlation coefficient ($[-1.0, 1.0]$) of two
random variables.

\begin{equation} \label{eq:corsnr}
  R_k(t) = \rho(\mathbf{H}(k), \mathbf{P}(t)) = \frac{\rho(\mathbf{H}(k), P_{exp})}{\sqrt{1 + \frac{1}{\mathit{SNR}}}}
\end{equation}

The $\mathit{SNR}$ is independent of the power model used by an attacker while
the $\rho(\mathbf{H}(k), P_{exp})$ is dependent on the power model, and it
describes how well an attacker approximates the $P_{exp}$. Equation
\refeq{eq:corsnr} gives several directions for defending correlation-based
attacks, and the goal is to make $R_k(t)$ approaching zero so that adversaries
do not easily get conclusive predictions.  Apparently a weak correlation
between $\mathbf{H}(k)$ and $P_{exp}$, or strong noises both reduce the
amplitude of $R_k(t)$, and consequently it becomes difficult for an adversary
to predict the right key. Usually, a lot more power trace measurements are
needed in case of a low $\mathit{SNR}$. Their exact relations still depend on
practical details.

Standaert \etal \cite{standaert2009unified} developed improved security metrics
using some information theoretic concepts. A few popular variations have been adopted
by the DPA Contest\footnote{www.dpacontest.org} which is an online benchmark
for power side-channel attacks. In the Non-Invasive Attack Testing Workshop
host by NIST in 2011, Goodwill \etal \cite{gilbert2011testing} introduced a
leakage detection method called the \textit{T-Test}. It uses a statistical
hypothesis testing method to detect if one or some sensitive intermediate
values can influence the measurement data. Hardware countermeasure evaluations
have not been widely using the new metrics and so this survey does not discuss
them in detail. Subsequent sections still use the number of power traces
because they appear the most commonly in publications. However, unless the
numbers are from the same authors and the same measurement setup, they should
not be considered as a precise metric for the effectiveness of corresponding
countermeasures.

\section{Overview of Countermeasures}

Ideally, making the cryptographic device physically secure or frequently
changing keys should be the most effective countermeasures for side-channel
attacks. However, such mitigations are difficult to guarantee in a large number
of cases. For example, an encrypted secret is shipped to an adversarial user
with the decryption key hardened in the device (\eg an encrypted FPGA design).
Or a strong attack already reveals the key before its session expires (\eg an
authentication smart card). Aggressive physical shielding is often difficult for
compact devices to adopt. Therefore, it is strongly desired for the
cryptographic circuit to have intrinsic fail-safe solutions in case it is in
the hands of adversaries.

Power side-channel attacks work because the power consumption of a
cryptographic device is dependent on its processed intermediate values, as
indicated by the $P_{exp}$. Such data dependencies are amplified by the
non-linear functions of modern ciphers, for example, the AES S-Box. The S-Box
was designed to resist conventional \textit{linear cryptanalysis}. But
experimental results of Guilley \etal in \cite{guilley2004differential} showed
the paradox that the better protected against linear cryptographic attacks a
block cipher is, the more vulnerable it is to differential power analysis.
Prouff also formally proved in \cite{prouff2005dpa} the resistance of an S-Box
to DPA attacks and the classical cryptographic criteria, such as high
non-linearity, cannot be satisfied simultaneously.

Hence, one obvious goal of counteracting techniques is to eliminate such
power-data dependencies, especially around the non-linear cipher functions. One
idea is to perform computations in a way such that different data processing
consumes the equal amount of power. Such techniques are known as
\textit{hiding}, which brings down $Var(P_{exp})$ ($\mathit{SNR} \to 0$). Thus
for all key hypotheses the correlation coefficients $R_k(t)$ are close to zero
and not conclusive. The capability to balance power consumptions is limited for
software. Hardware design can effectively modify a chip's power characteristics
at various levels.

In practice, the data dependency can not be completely removed, and then
according to the $\mathit{SNR}$ definition another direction for mitigation is to
increase the variance of the \textit{amplitude noise}. Noise can be achieved by
\textit{on-chip noise generation}, including excessive switching noises and
electronic noises. Hardware implementations using wide data paths (\eg parallel
S-Boxes) for cryptographic algorithms are example approaches to increase
switching noises. The added noises also do not completely hide the exploitable
power component, but they can fairly complicate an attack. Programmers can
increase noise by executing parallel activities to the cryptographic functions.
But doing multiple tasks in software is often limited by the cost and footprint
constraints on many security ICs.

\textit{Masking}, also known as \textit{secret sharing}, is a general
countermeasure for both software and hardware that randomizes the intermediate
values while still keeps the input and output of a cipher the same as the
unmasked version. For example, in the AES encryption algorithm, the plaintext
and the key are added or multiplied with random values, \ie the masks. The
transformations of masks through round functions are tracked, and before the
ciphertext output the masking effect is removed properly. Consequently, it
becomes difficult for adversaries to correlate the $P_{exp}$ using hypothetical
power models ($\rho(\mathbf{H}(k), P_{exp}) \to 0$). Masking techniques can
apply to the algorithmic level without changing the power consumption
characteristics of the cryptographic hardware, and therefore it is very
commonly used in software. The same algorithmic masking can be ported to
hardware, and besides algorithmic masking, hardware design also utilizes
gate-level masking. Because masking is suitable for multiple design levels,
there is a large body of research literature on this topic. This survey will
focus on gate-level masking in the next section.

Another technique that is implicit in the $\mathit{SNR}$ expression is to break
the trace alignment effectively (see \refsec{sec:signal-processing}), and
therefore it prevents attackers from building a meaningful power trace matrix.
The software can implement this countermeasure can be implemented by inserting
random number of \textit{dummy instructions} before and after protected
functions, or \textit{shuffling} the execution order of S-Boxes for every data
block.  Hardware design can adopt similar techniques. Effects that cause trace
misalignment is also called \textit{temporal noise} for attacks.

Usually, a cipher implemented in hardware is more resistant to power analysis
attacks than its software version. Microprocessors are often highly
pipelined and equipped with a data bus that consumes the majority part of the
whole chip's dynamic power. Simple power models such as Hamming weight and
Hamming distance often fairly represent the exploitable power of their cipher
implementations. The parallel nature of hardware, complex data paths, and
irregular transistor toggles in one clock cycle all increase the attacking
effort. Mangard \etal \cite{mangard2008power} designed an AES core in ASIC
using four parallel S-Boxes and found it took significantly more effort to
attack the hardware than the software counterpart on an 8051 microcontroller.


Complex SoCs and devices may not be more resistant to power side-channel
attacks than a small chip. Successful side-channel attacks mounted to modern
smartphone processors\footnote{These two cases actually use electromagnetic
signals for attacking but the principle of leakage is similar to the power
side-channel, both induced by rapidly changing currents.} are demonstrated by
\cite{jun2012your} and \cite{nakano2014pre}. A large design usually has many
power and ground pins surrounding the chip. Some of the supply pins are close
or dedicated to a security circuit, and thus a quality measurement can be done
with these pins. Even worse, for power efficiency and hence usability,
contemporary large-scale chips must use power gating techniques to shut down
unwanted switching of transistors. From a security perspective these techniques
effectively turn off switching noises and leave the running cryptographic
module as vulnerable as in a small chip. Therefore appropriate countermeasures
must also be employed cryptographic circuits of large-scale SoCs as well.




\clearpage
\section{Hardware Countermeasures}

Hardware design can effectively implement a variety of countermeasures. Since
power consumption properties can be largely changed in hardware design, many
approaches have been trying to conceal the real activities of the protected
circuit. This section first introduces logic gate masking, followed by the
dual-rail precharge logic that intends to equalize power consumption for all
switching activities. Gate masking and dual-rail/precharge designs often
combine to take merits of both. Other techniques such as power line isolation,
noise generation and randomized dynamic voltage and frequency scaling have also
been proposed to thwart DPA attacks.


\subsection{Masked Logic}

\begin{figure}[h]
\centering
  \includegraphics[width=.45\textwidth]{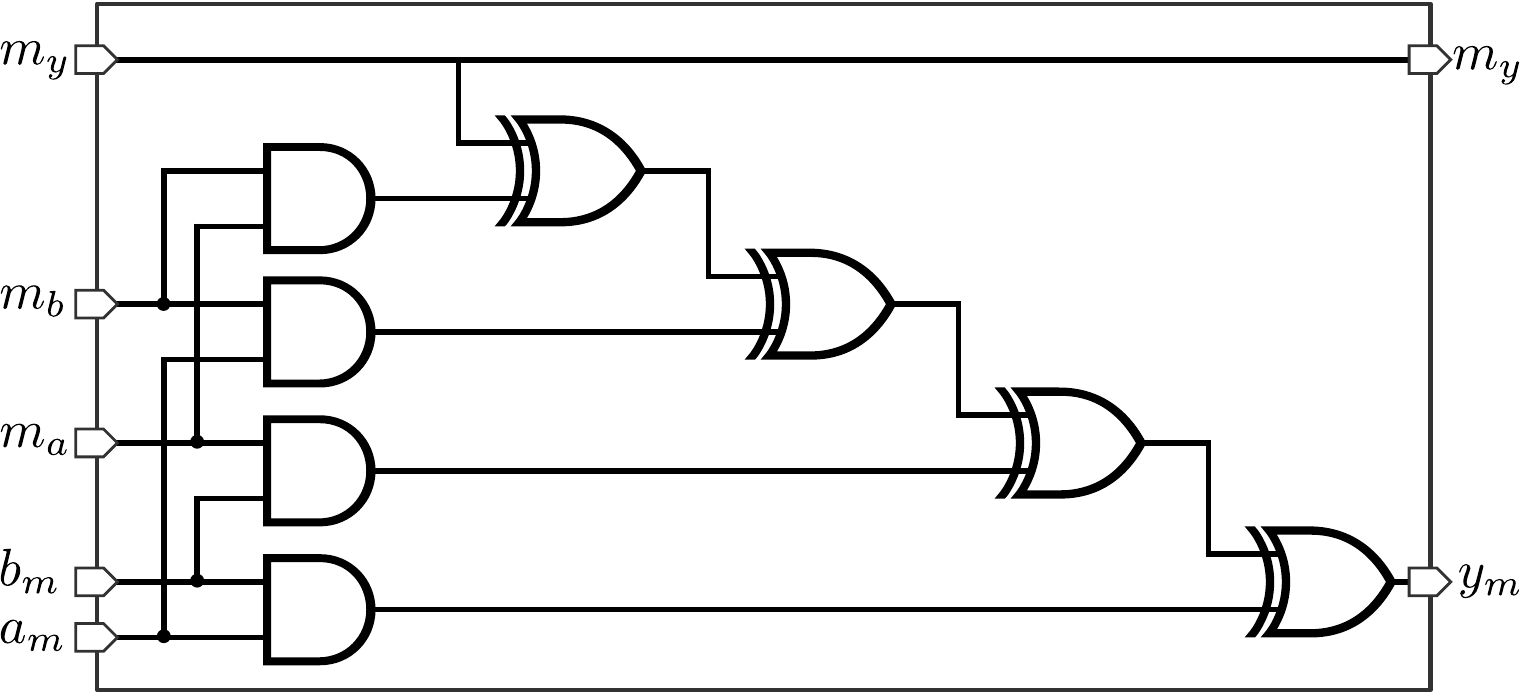}
  \caption{The Masked \texttt{AND} Gate}
  \label{fig:maskedand}
\end{figure}

Alternative to masking at algorithmic level is the use of \textit{masked logic gates}
so that at the gate-level no values stored in hardware are correlated to
secret intermediate values. The basic idea of gate masking is to mask properly
a gate's input and output signals and never let the true secret values expose
until the final output. For example, a regular 2-input \texttt{AND} gate ($y = a \cdot b$)
is transformed to a 5-input ($a_m$, $m_a$, $b_m$, $m_b$, $m_y$), 2-output ($y_m$, $m_y$)
\texttt{Masked-AND} \cite{trichina2003combinational} as shown in
\reffig{fig:maskedand} and equation \refeq{eq:maskedand}. The $m_a$, $m_b$, and
$m_y$ are random 1-bit masks that can be either equal or different. In the
same manner, other basic gates can be converted into masked versions and
consequently any circuit can be built using masked gates. Ishai \etal
\cite{ishai2003private} formally proved that any logic circuit can
transform into corresponding securely masked versions. Before entering the
masked circuit inputs like $a$ and $b$ are masked as $a_m = a \oplus m_a$ and
$b_m = b \oplus m_b$ (remember that \texttt{XOR} functions are not easily
attacked by DPA). Masked values and mask bits propagate together along the
circuitry (\eg one AES round) and at the end of the protected computing the
true values are recovered by removing masks (\eg $y = y_m \oplus m_y$).




\begin{equation} \label{eq:maskedand}
\begin{split}
  y_m = & ~ a_m \cdot b_m \oplus (m_a \cdot b_m \oplus (m_b \cdot a_m \oplus (m_a \cdot m_b \oplus m_y))) \\
        & with: ~ a_m = a \oplus m_a, b_m = b \oplus m_b, y_m = y \oplus m_y
\end{split}
\end{equation}

Compared to algorithmic masking, masking at gate level does not have to
consider the higher level functions of using logic gates. Given a cryptographic
circuit implemented by regular gates, masked gates can be swapped for them to form
a less leaky version, at a cost of increased gate count and extra power
consumption. Standard logic synthesis tools do not well support this design flow,
whose primary goals are usually optimizing timing and area. Therefore,
additional effort is required for the functional and timing verification of the
masked gate netlist.

Masking is a mathematically provable securing scheme without considering real
hardware behaviors. It effectively hides dependencies between the secret data
and \textit{settled} final values of gate outputs, and counteracts basic power
models such as the Hamming weight and the Hamming distance because such attacks
also just consider the final values. Unfortunately, the real hardware is
complicated, and a masked design often overlooks glitches caused by the
different arrival times of input signals to logic gates. It is obvious that the
masked gate in \reffig{fig:maskedand} has severely unbalanced signal paths,
and worse variations may already exist before they enter the first level of the
masked layers. Consequently some unwanted glitches appear at the output before
the final value settles, and they are found correlated to the secret key
values. Mangard \etal \cite{mangard2005side} demonstrated that simply masked
gates are vulnerable to attacks using the more precise toggle count power
model. A formal explanation of how glitches leak information is given by Nikiva
\etal \cite{nikova2006threshold}.

Masking methods, including algorithmic masks, are also susceptible to
various advanced DPA attacks, such as higher-order DPA
(\cite{messerges2000using}), template attack (\cite{agrawal2005templates}), and
algebraic side-channel attack (\cite{renauld2010algebraic}.  These attacks
exploit multiple intermediate values in a power trace and cross-correlate them
to remove the effect of masks. On the other hand higher-order masking methods
(\cite{rivain2010provably}) have been proposed to counteract higher-order power
analysis. The \textit{masking order} is determined by how many independent data
\textit{shares} are needed to represent the true secret value. If the number of
shares is $d + 1$, \eg $X = \bigoplus_{i = 1}^{d + 1} x_i$, the scheme is known
as the $d^\mathit{th}$-order masking. For example a masked gate's output $y$ is
split into two shares, \ie $y_m$ and $m_y$; this is called a first-order
masking. A $d^\mathit{th}$-order masking is supposed to resist a
$d^\mathit{th}$-order attack. Higher-order masking is difficult for efficiently
hardware implementations and flaws \cite{coron2014higher} are also found due to
over complex designs. Although simple masking is often not enough to prevent
successful attacks, it certainly increases the difficulty level and usually
achieves better resistance when combined with other countermeasures.


\subsection{Dual-Rail Precharge (DRP) Logic}

The DRP logic cells are designed to make output switching activities
independent of inputs to achieve constant power consumptions at the gate level.
In a dual-rail gate, logic \texttt{0} and \texttt{1} are represented by
complementary pairs \texttt{(0, 1)} and \texttt{(1, 0)} respectively. The
precharge state is defined by \texttt{(0, 0)}, but \texttt{(1, 1)} is not a
valid circuit state. The promise is exactly the same count of \texttt{0} $\to$
\texttt{1} and \texttt{1} $\to$ \texttt{0} transitions appear per gate and per
clock cycle, regardless of the input activity. The last four columns of
\reftab{tab:switching} describes switching activities of a dynamic
\texttt{(AND, NAND)} pair. The shaded transitions occur when a
\textit{precharge} signal asserts and the unshaded ones happen when the it
de-asserts. In dynamic CMOS logic design, the second step is the
\textit{evaluation} phase.

\begin{table}[h]
\centering
  \includegraphics[width=.39\textwidth]{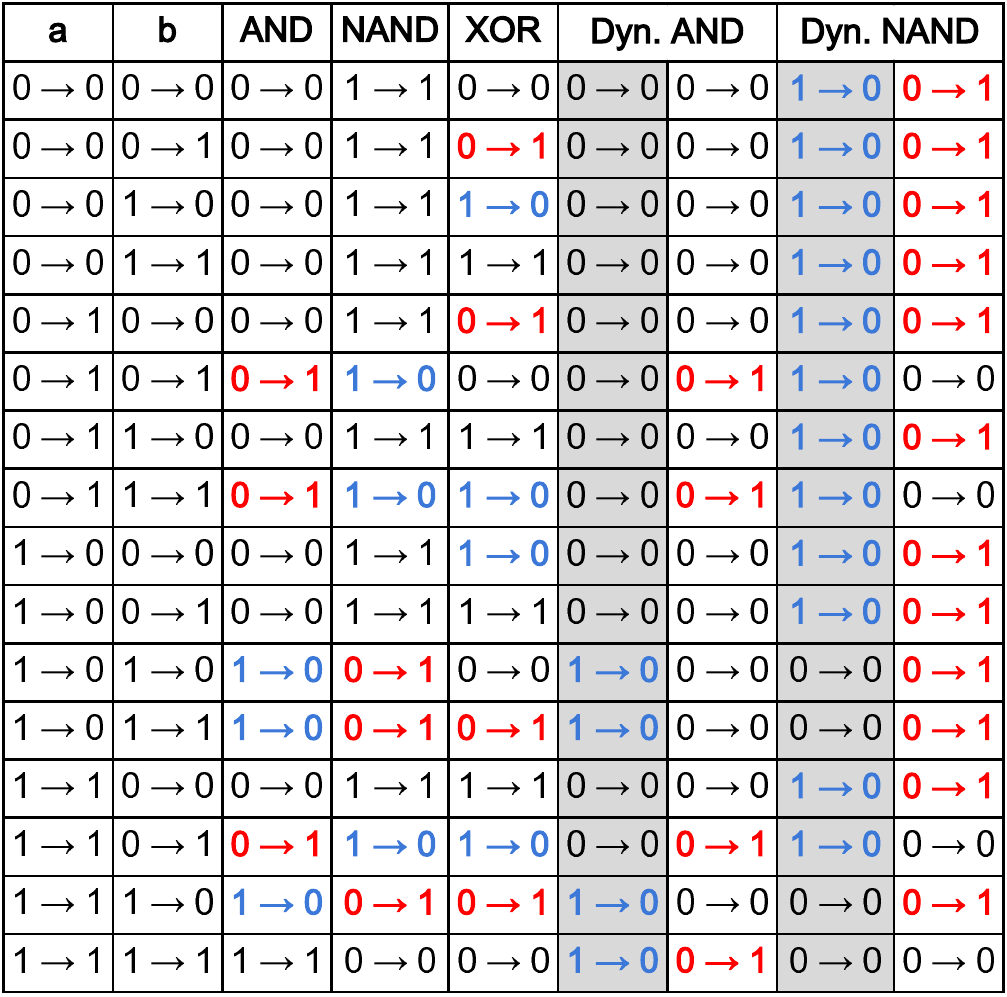}
  \caption{Switching Activities of Gates. For dynamic gates the shaded column indicates the precharge phase.}
  \label{tab:switching}
\end{table}

\begin{figure}
\centering
  \includegraphics[width=.39\textwidth]{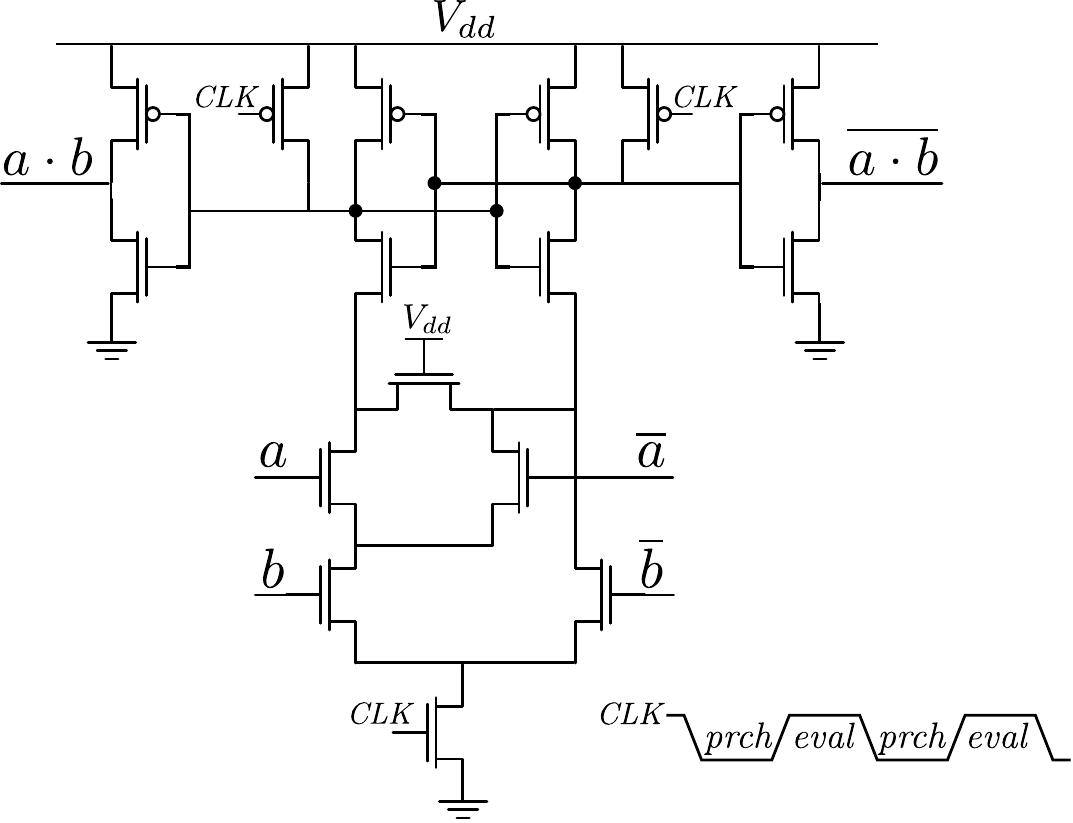}
  \caption{The SABL \texttt{AND} Gate}
  \label{fig:sabl}
\end{figure}

Classical dual-rail domino logic gates satisfy the desired pattern of
switching. But one essential design metric is to ensure all charging and
discharging paths have the same capacitance. Tiri \etal \cite{tiri2002dynamic}
reported the earliest design of DRP cells in academia for counteracting DPA
attacks. On the basis of (modified) dual-rail domino logic, their solution is
to use a pair of cross-coupled inverters for capacitance balancing, shown in
\reffig{fig:sabl}. Since the cross-coupled inverters are also known as a sense
amplifier, this transistor-level DRP cell design is called \textbf{Sense
Amplifier Based Logic (SABL)}. An SABL cell has the following advantages: 1)
outputs switching activities independent of inputs, 2) immunity to glitches,
and 3) balanced internal capacitances. The immunity to glitches is because
during the precharge phase all outputs only monotonically fall, and in the
evaluation phase the outputs can only monotonically rise (nature of dynamic
logic). However, a design based on SABL cells still has to balance external
loads and especially the parasitics on routing wires. To relax the routing
constraint, a \textit{Three-Phase Dual-Rail Pre-Charge Logic (TDPL)}
\cite{bucci2006three} was developed, by enforcing an additional
discharge\footnote{Depending on different nodes for circuit analysis, the
precharge or discharge could lead to either \texttt{0} or \texttt{1}. Looking
at the outputs of gates, precharge drops all outputs to \texttt{0} but
discharge brings them to \texttt{1}.} phase after the evaluation. Consequently
in TDPL either differential output exactly charges once and discharges once.
As a result, unbalanced external load capacitance has a much lower effect.

\begin{figure}[h]
  \centering
  \includegraphics[width=0.21\textwidth]{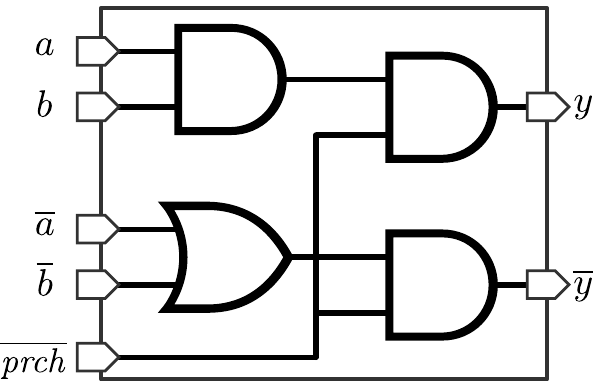}
  \caption{The SDDL \texttt{AND} Gate}
  \label{fig:sddl}
\end{figure}

The SABL design does achieve nearly constant power consumption in circuit
simulation but it requires a tedious and expensive full-custom IC design flow,
especially for its dynamic logic. Consequently, many later developments of DRP
cells intend to use static CMOS standard cells to emulate the switching
activities of SABL cells. Tiri \etal \cite{tiri2004logic} also introduced the
\textbf{Wave Dynamic Differential Logic (WDDL)} that is compatible with
standard cell based semi-custom ASIC design. The authors first demonstrated a
\textit{Simple Dynamic Differential Logic (SDDL)} built by basic static CMOS
gates of a standard cell library to emulate the dual-rail precharge behavior.
An SDDL \texttt{AND} gate is illustrated in \reffig{fig:sddl}. The dual gate to
the \texttt{AND}, the \texttt{OR} gate, is derived using the De Morgan's law:
$\overline{a \cdot b} = \overline{a} + \overline{b}$; and their outputs are
\texttt{AND}-ed with the precharge signal. To avoid glitching, the
dual-rail functions must implement positive monotonic gates
(outputs always swing in the same direction as inputs). The original
publication \cite{tiri2004logic} only allows the \texttt{AND} and the
\texttt{OR} gates\footnote{Any boolean logic can be implemented with the
\texttt{AND}, \texttt{OR}, and \texttt{INV} operators.}. Inversion is inherent
in a dual-rail logic gate by swapping the two outputs.


\begin{figure}[h]
  \centering
  \includegraphics[width=0.45\textwidth]{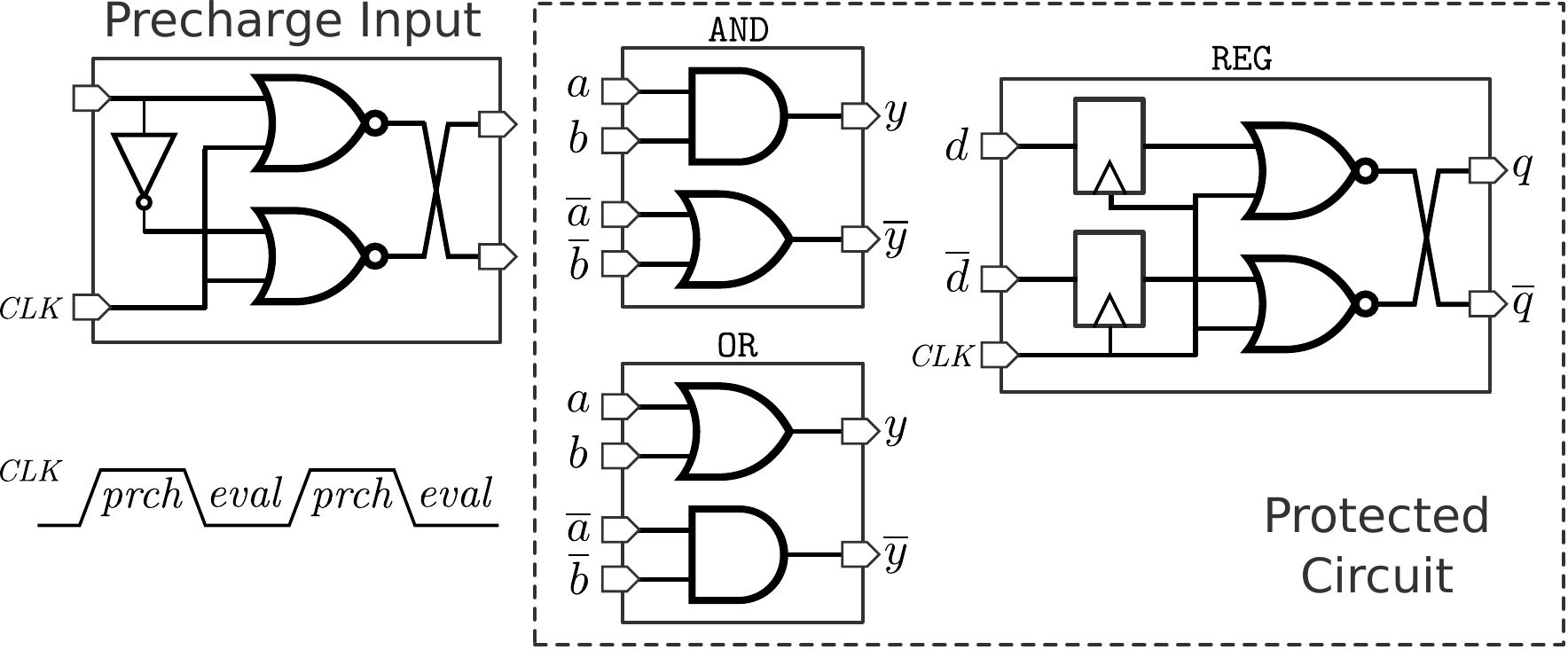}
  \caption{The WDDL Encryption Module Design}
  \label{fig:wddl}
\end{figure}


By virtue of the positive monotonic gates, the precharge signal does not have
to be globally distributed to every combinational gate. During the precharge
phase, the inputs of any compound gate are precharged to logic \texttt{0}s and
any \texttt{AND} or \texttt{OR} gate outputs logic \texttt{0} as well.
Accordingly, the two \texttt{AND} gates at the output stage of an SDDL gate are
removed, and the precharge signal (logic \texttt{0}) can propagate like a wave
to all combinational gates and stop at clocked flip-flops. Hence comes the name
WDDL.  The precharge waves are launched from register cells as shown in
\reffig{fig:wddl}. Inverting gates are not allowed along the way because they
halt the \texttt{0}-waves. A WDDL gate has the same switching activity as in
SABL. But one good property that WDDL loses is the balanced internal
capacitance: the \texttt{AND} and \texttt{OR} gates that compose a WDDL gate
are not equal. Therefore, in general not all internal nodes of a WDDL gate are
charged and discharged in the same way. The authors argue that with technology
shrinking the interconnection capacitance is more dominant, and the focus
should be on routing balance.


Although a WDDL design can be implemented using existing standard cells, normal
IC design tools still do not natively support it. Suppose a cryptographic
circuit is designed by hardware description languages (Verilog or VHDL), the
designer first has to restrict the use of target cells to only \texttt{AND},
\texttt{OR} and \texttt{INV}. Once have the synthesized cell netlist, a
custom script is executed to replace each gate by its WDDL counterpart and
remove inverters by exchanging the outputs. From the gate netlist to the physical
layout strict design constraints must be applied, especially for differential
wire routing. The constraints for routing the differential wires include but
are not limited to the follows \cite{tiri2006digital}: 1) always on the same
metal layers, 2) always on adjacent routing tracks, 3) always have the same
lengths and widths, 4) always have the same number of vias for resistance
balancing, 5) must be shielded with $V_{dd}$ or ground lines to remove
crosstalk, 6) making every other metal layer a ground plane to control the
inter-layer capacitance. Therefore, the designers have to force commercial IC
design tools in many ways to meet the physical design constraints. In each
design iteration, parasitic capacitance analysis must be rigorous performed.
Moreover, due to custom editing of the gate netlist, formal verification and
timing closure require greater effort than in a normal IC design. For FPGAs,
the constraints are even harder to meet than the ASIC case due to the limited
degree of freedom in the FPGA physical design.

An SoC\footnote{The authors claim this is the first practical DPA
countermeasure implemented and tested in actual silicon.}
\cite{tiri2006digital} using WDDL design as its AES portion was fabricated using
a TSMC 0.18-$\mu$m process. On the same chip, an insecure AES counterpart exists
as well, comprising regular standard cells and regular physical design rules.
Both parts have been implemented starting from the same synthesized gate
netlist. In the evaluation, for the insecure AES core on average 2,000 power traces
are required to disclose a key byte. The AES in WDDL successfully protects 5
out of 16 key bytes up to 1,500,000 power traces; the 11 leaked bytes on
average needs 255,000 traces each. The increased security is at a cost of 3x
increase in area, 4x increase in power consumption and a limited operating
speed up to 125MHz. The residual leakage of this WDDL design was due to imperfect
routing and the lack of IC design tool support on security.

In theory, the power consumption of an SABL or WDDL gate would be independent of
the data values processed and so thwart power analysis attacks. However in
practical IC design the capacitances of the complementary outputs are
impossible to be perfectly balanced due to all the complications in logic
synthesis, placement, and routing. Even if the most powerful IC design tool can
make such a balance at design time when the chips are being fabricated there
are still manufacture process variations that effectively break the balance.
Consequently, there are always residual leakages and given sufficient number of
power traces the secrets can still be extracted. A DES cryptoprocessor
implemented in FPGA using WDDL without place and route constraints was
successfully attacked by \cite{sauvage2009successful}.



\subsection{Masked Dual-Rail/Precharge Logic} \label{sec:rsl_mdpl}
The strong constraints in IC physical design for DRP logic cells motivate
research for more efficient solutions. Suzuki \etal \cite{suzuki2004random}
proposed the \textbf{Random Switching Logic (RSL)} that combines the mask and
the precharge, but it uses single-rail logic. It introduces the precharge
signal for partially resisting glitches, and the random switching bit (mask) is
employed to avoid dual-rail and routing balance constraints. An RSL gate
satisfies two properties: 1) executes masked operations for all input/output
signals using the same 1-bit random mask, 2) executes operations while the
precharge (like enable) is logic \texttt{1}; otherwise the output drives
\texttt{0}. For example a 2-input \texttt{NAND/NOR} gate in RSL is expressed
by equation \refeq{eq:rsl}. The $e$ signal is for precharge, and the $m$ is a
1-bit mask that switches the two functions randomly. When $m =$ \texttt{0}, the
\texttt{RSL-NAND/NOR} works as a regular \texttt{NAND} and when $m =$
\texttt{1} it works as a regular \texttt{NOR}. To encrypt a data byte, the $m$
bit is randomly generated and expanded to a byte, \texttt{XOR}-ed with the
input data byte, the round key byte, fed into the protected circuit to switch
logic roles, and in the end \texttt{XOR}-ed with the masked output to get the
real encrypted value. Because the RSL gates do not confine to positive
monotonic functions, the design is not inherently free of glitches. To ensure
glitch-free, the precharge signals must de-assert after all input data signals
settle. Suzuki \etal claims this can be achieved easily even by automatic
place-and-route tools if enough delay time is given.


\begin{equation} \label{eq:rsl}
\begin{split}
  & \mathtt{RSL ~ NAND/NOR}: ~ y_m = \overline{\overline{e} + a_m \cdot b_m + (a_m + b_m) \cdot m} \\
  & ~~~~ with: ~ a_m = a \oplus m, b_m = b \oplus m, y_m = y \oplus m
\end{split}
\end{equation}

RSL gates do not have uniform switching activity as shown in
\reftab{tab:switching}, but it intends to hide the data induced variations by
randomness. The authors of RSL evaluated their design in FPGA and compared with
a WDDL design\footnote{The WDDL design was implemented in the same FPGA, but
the authors did not mention any control in balancing the WDDL loads.}, the RSL
needs 1/3 of the area of WDDL with better timing performance. Under DoM based
attacks, the WDDL design leaks with 60,000 traces but the RSL resists 200,000
queries.

\begin{figure}
\centering
  \begin{subfigure}[b]{0.23\textwidth}
    \includegraphics[width=\textwidth]{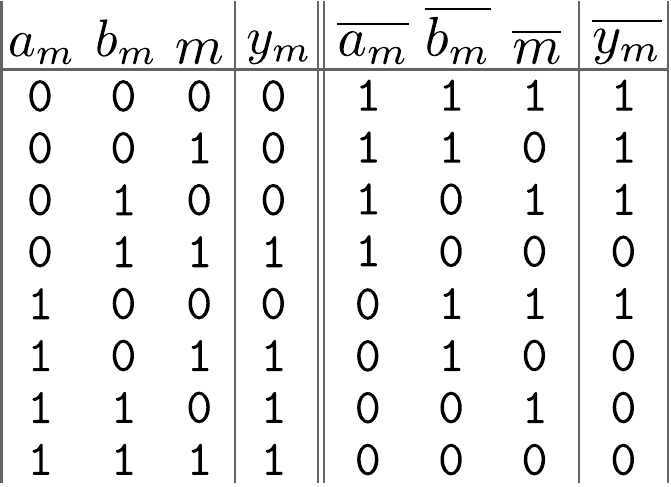}
    \caption{} \label{fig:mdpland}
  \end{subfigure}
  \hfill
  \begin{subfigure}[b]{0.20\textwidth}
    \includegraphics[width=\textwidth]{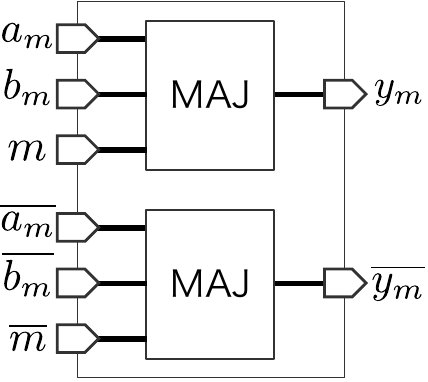}
    \caption{} \label{fig:mdplmaj}
  \end{subfigure}
  \caption{(a): The MDPL \texttt{AND} Gate Truth Table
           (b): The MDPL \texttt{AND} Gate Schematic}
\end{figure}

\textbf{Masked Dual-Rail Precharge Logic (MDPL)} \cite{popp2005masked}, as its
name explains, intends to combine merits of both mask and DRP to halt DPA
attacks and not pose any tedious IC design constraints. The MDPL adopts
dual-rail design to have data-independent output switching activity and
intends to use a random mask bit to compensate for the loss of routing balance.
An MDPL \texttt{AND} gate takes six dual-rail inputs
($a_m$, $\overline{a_m}$, $b_m$, $\overline{b_m}$, $m$, $\overline{m}$)
and produces two outputs ($y_m$, $\overline{y_m}$). The truth table is in
\reffig{fig:mdpland} and the outputs are computed as
$y_m = ((a_m \oplus m) \cdot (b_m \oplus m)) \oplus m$, and
$\overline{y_m} = ((\overline{a_m} \oplus \overline{m}) \cdot (\overline{b_m} \oplus \overline{m})) \oplus \overline{m}$
(just for the truth table, the implementation should not be done this way).
From the truth table, it is observed that the $y_m$ and $\overline{y_m}$ can
be computed by the \textit{majority} function (\textit{MAJ}) which
outputs \texttt{1} if the inputs have more \texttt{1}s than \texttt{0}s, or it
outputs \texttt{0}. A gate that performs the majority function is supposed to
exist in a standard cell library. In MDPL, all signals are precharged to
\texttt{0} and similar to WDDL the precharge waves are launched from registers
and sweep all combinational logic. Since the majority gate is a positive
monotonic function, the MDPL gates are immune to glitches. Moreover, because
both the dual-rail outputs can be computed by the same majority function, MDPL
has internally balanced current paths.


Like RSL, MDPL also uses a single-bit mask $m$ (and its complementary
$\overline{m}$) for each data byte encryption. MDPL does not require balanced
loads and routing constraints. The imbalance is supposedly compensated by the
random masks. The authors simulated their design using a 0.35-$\mu$m standard
cell library, evaluated that by mounting DoM attacks on the gate-level netlist
power simulation and concluded the secret key was not discovered. Compared to a
regular but leaky standard cell design, the MDPL version is 4.5x bigger in the
area and half the operating speed.

Although SABL, WDDL, RSL and MDPL are all designed to be glitch-free, their
implementation still can not prevent different arrival times of signals that
depend on cell and route delays. Many of the normal boolean logic gates have
the property that their logical outputs can be uniquely determined without
necessarily knowing all the inputs. Accordingly a gate sometimes evaluates its
logical output early, before waiting for all of its logical inputs (not a
timing violation). This shifted time of evaluation, called \textit{early
propagation} (or \textit{early evaluation}), causes data-dependent power
dynamics that are exploitable. Suzuki and Saeki \cite{suzuki2006security}
explained the leakage using WDDL and MDPL as examples. They pointed that a cell
that avoids early propagation must delay the evaluation moment until all input
signals have arrived, \eg all inputs of MDPL have settled to differential
values. Popp \etal \cite{popp2007evaluation} confirmed the leakage on a
prototype MDPL chip and proposed improved MDPL (iMDPL) to counteract early
propagations. The remedy in iMDPL significantly increased the area compared to
MDPL. For example, one more \texttt{NAND}, three more \texttt{OR} and six more
\texttt{NOR} gates were added to the MDPL \texttt{AND} gate in
\reffig{fig:mdplmaj}.


Tiri and Schaumont \cite{tiri2007changing} found another source of
leakage in RSL and MDPL that is the mask bit itself. For example in RSL, the
encrypting process with $m =$ \texttt{0} consumes more (or less)
power than the encrypting process with $m =$ \texttt{1}. This is expected as
the mask bit puts the circuit in one of the two complementary forms. In other
words, the distribution of samples in some columns of the power trace matrix
$\mathbf{P}$ already looks like \reffig{fig:gaussian2} with an observable notch
in the middle. Finding the time when the two distributions clearly separate
deserves some searching effort. Tiri and Schaumont \cite{schaumont2007masking}
suggest folding one part around the notch on top of the other and then perform
a regular DoM based attack. Therefore, later research work often refers to this
technique as the \textit{folding attack}. The authors used toggle count
simulation and proved the concept. Because the Gaussian curve is a
probability density function, using it to filter the mask bit is called
\textit{probability density function filtering}. For MDPL, a similar approach
is applicable too. Practical folding attacks on MDPL were reported by De Mulder
\etal \cite{de2009practical} on a prototype chip. Moradi \etal
\cite{moradi2012masked} further found leakage in iMDPL using the folding attack
and a few other advanced mask detections methods. So once the mask effect
disappears from RSL or MDPL countermeasures, the circuit downgrades to simple
precharge or dual-rail logic without balanced load capacitance and is
susceptible to attacks.



\subsection{Power Line Isolation}

Alternative to manipulating the logic gates, another approach intends to
isolate/decouple the cryptographic circuit's power supply from the external
voltage source. Shamir \cite{shamir2000protecting} introduced a power line
decoupling method using two capacitors that work in an alternating manner under
a proper switch control. During half the time, one capacitor is charged by the
external power source while the other capacitor is discharged by supplying the
security chip. During the other half of time, the two capacitors switch roles.
Because the charges stored in the supplying capacitor can only sustain a
limited number of circuit operations, this paper mainly discussed the
feasibility of using a large capacitive but separate circuit module for
securing the protected chip. Both can fit in the cavity of a smart card.

Since externally placed switch capacitors are easily tampered by manipulating
their pins, later developments focus on pushing them into the same chip.
Corsonello \etal \cite{corsonello2006integrated} proposed an integrated
\textit{charge pump} also using switches and capacitors. Although not evaluated
under practical attacks, the authors claimed a strong effectiveness of their
design based on simulation results.

\begin{figure}
\centering
  \begin{subfigure}[b]{0.25\textwidth}
    \includegraphics[width=\textwidth]{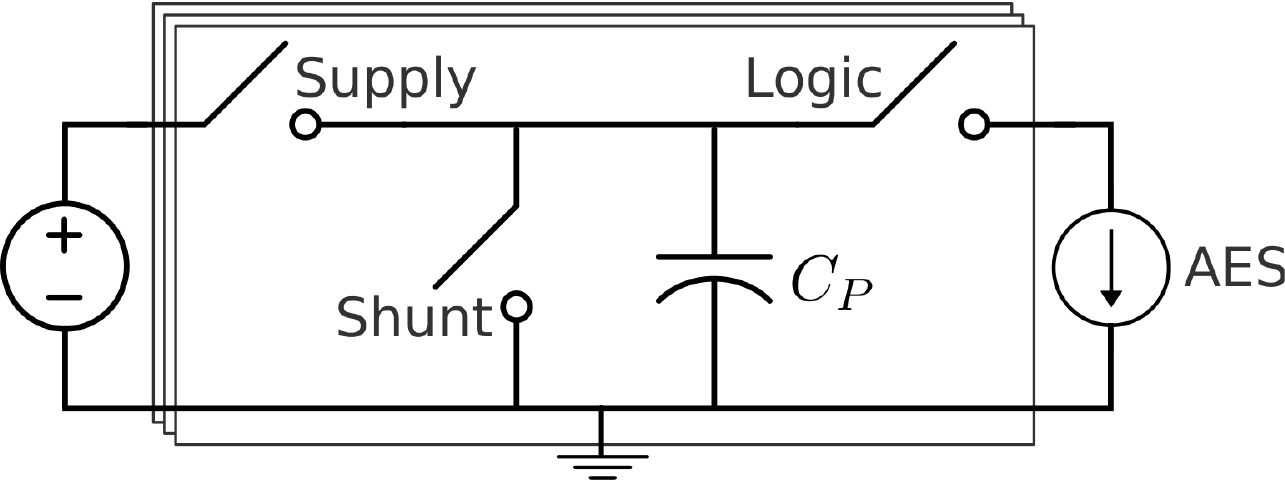}
    \caption{} \label{fig:sclblock}
  \end{subfigure}
  \begin{subfigure}[b]{0.21\textwidth}
    \includegraphics[width=\textwidth]{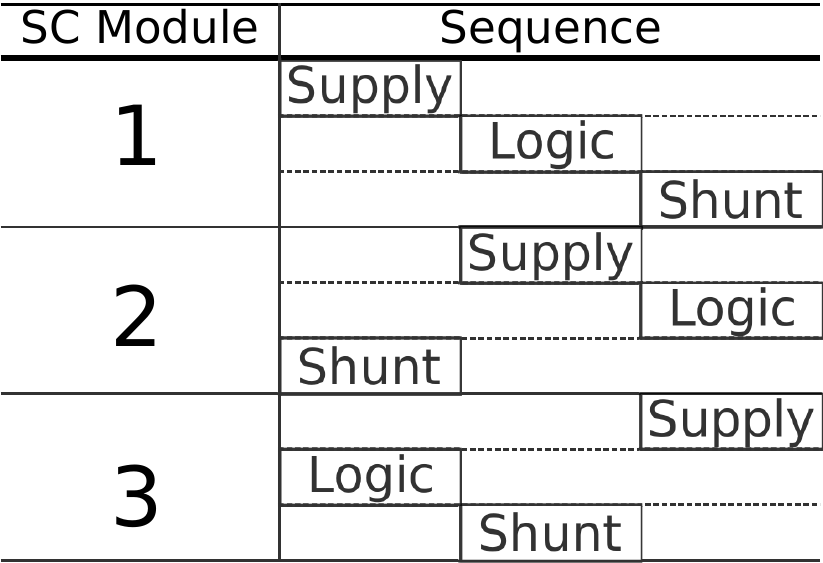}
    \caption{} \label{fig:sclclock}
  \end{subfigure}
  \caption{(a): There Switching Capacitor Modules of \cite{tokunaga2010securing}
           (b): The Switching Pattern Diagram. Named blocks indicate conducting of corresponding switches.}
\end{figure}

Tokunaga and Blaauw \cite{tokunaga2010securing} introduced a switching
capacitor based current equalizer and practically evaluated that in actual
silicon. \reffig{fig:sclblock} shows the design consisting three switching
capacitor modules, each having a 100-pF capacitor $C_P$ for charge storage and
three controlled pass-transistor switches namely $S_{\mathit{Supply}}$,
$S_{\mathit{Logic}}$, and $S_{\mathit{Shunt}}$. Each capacitor module has three
cyclic switching states: 1) Only $S_{\mathit{Supply}}$ conducts and replenishes
charge from the external supply, 2) Only $S_{\mathit{Logic}}$ is closed to
provide charge for the protected circuit, and 3) Only $S_{\mathit{Shunt}}$
conducts and provides a path for $C_P$ to discharge. The three modules have
staggered switching pattern, shown in \reffig{fig:sclclock}, to ensure
uninterrupted operation of the protected circuit. The storage capacitor was
implemented using a combination of NMOS and metal-to-metal capacitors. During
a discharge phase, its voltage is monitored by a comparator, and when its
voltage discharges to a preset constant $V_{\mathit{Ref}}$, a trigger will
open $S_{\mathit{Shunt}}$ to stop discharging. Thus $V_{\mathit{Ref}}$ is to
ensure no remainder variable voltage is detectable at the beginning of the next
charging phase, and it prevents discharging to zero to save energy.

The test chip fabricated using a 0.13-$\mu$m CMOS process has a switching
capacitor protected AES core as well as a unprotected one for practical
comparison. The chip design flow allows the AES core placed and routed
using standard CAD tools, and the current equalizer module is integrated
after that. Before fabrication, some mixed-signal circuit simulation was
performed to verify the correct operations. The protected AES circuit did not
leak any secret during CPA attacks up to 10 million power traces while the
insecure AES core breaks at 10,000 measurements. The switching capacitor block
incurs a 7.2\% increase in the area, 33\% increase in the power consumption,
and half the performance loss.

\begin{figure}
  \centering
  \includegraphics[width=.39\textwidth]{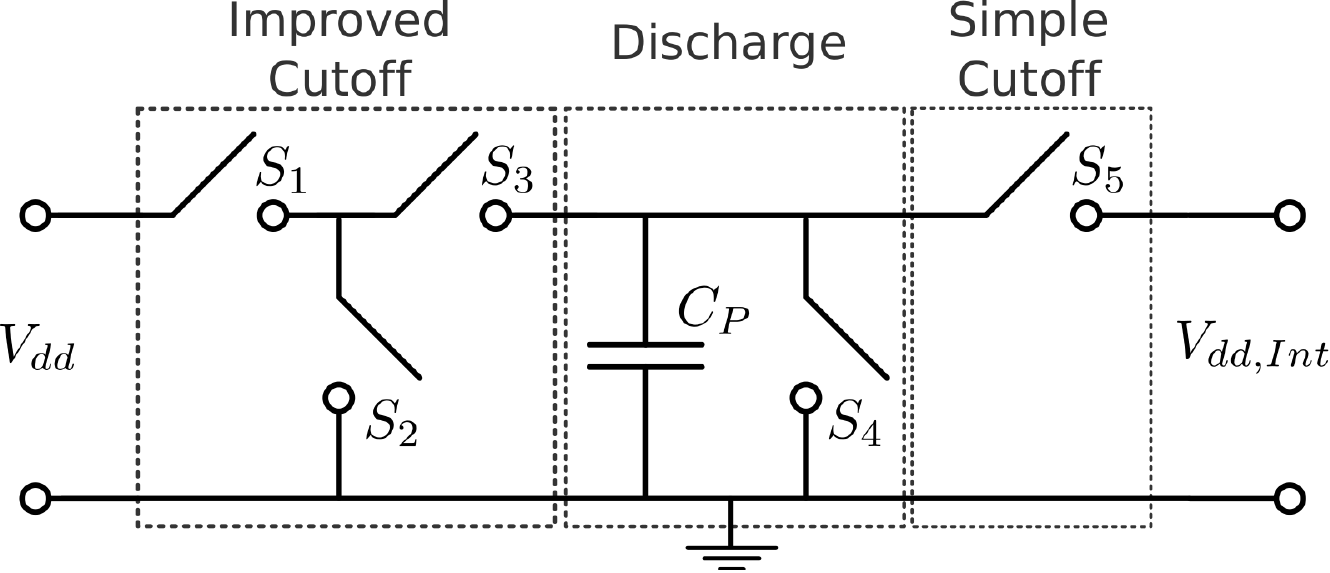}
  \caption{The Decoupling Circuit of \cite{gornikhardware}}
  \label{fig:scl15}
\end{figure}

While \cite{tokunaga2010securing} was designed as a bulk unit sitting aside the
protected circuit, Gornik \etal \cite{gornikhardware} bring switching
capacitors to individual logic gates. They pointed out two issues of
\cite{tokunaga2010securing}: 1) its effectiveness depends on the quality of
switches, which in semiconductor is not perfect and their sub-threshold
currents in cutoff mode is exploitable by adversaries. 2) the mixed-signal
design flow is hardly reused for other chips. To resolve the first problem,
Gornik \etal improved the external cutoff circuit (\reffig{fig:scl15}) to
reduce the impact of sub-threshold currents in switches. Compared to
\reffig{fig:sclblock}, the $S_1$ and $S_3$ switches work like the
$S_{\mathit{Supply}}$, the $S_4$ is equivalent to the $S_{\mathit{Shunt}}$, and
the $S_2$ and $S_5$ switches resemble the $S_{\mathit{Logic}}$. $S_2$ is the
improvement to accommodate sub-threshold currents of $S_1$ and $S_3$: when
$S_1$ and $S_3$ are open but leaking, $S_2$ should effectively lead the
currents to ground to avoid crosstalk between $V_{dd}$ and
$V_{\mathit{dd,Int}}$.

For effortless integration in digital IC design, the switching capacitors are
compressed into a new standard cell and can be distributed to each logic gate
to supply it for several transitions. The concept is to design the decoupling
cell once, and it can be reused and compatible with automated layout tools.
However, using their target 0.15-$\mu$m CMOS technology the decoupling cell was
259.78 $\mu m^2$, 10x larger than a 2-input \texttt{XOR} gate of the standard
cell library. Existing physical design tools did not support automated layout
of such decoupling cells, and so they performed manual placement and routing of
the test circuit which incurred only moderate effort due to its small size. The
authors anticipated the area overhead could be reduced using a smaller
technology node, and they could use automation tools for future chip layout.
The detailed transistor-level design of the cell is in \cite{gornikhardware}
but the paper does not disclose specifications for $C_P$, \eg how large
capacitance it should have.

%

Chips each having a protected S-Box of the PRESENT block cipher were fabricated
and evaluated. Under assessment, the batch of chips from the same design, same
wafer, and the same package exhibited quite diverse resistance to DPA attacks.
Some of them showed a high level of robustness while some of them provided nearly
no protection. The authors analyzed their chips, and they thought the reason was
the quality of chip bondings. They received the manufactured dies
unpackaged and performed packaging and wedge bonding themselves. By repeating
the bonding process for the same test chip, they confirmed different results
using the same setup for power line attacks. The bonding issue was expected to
be resolved using advanced bonding machines.

Two things the recent publications did not precisely discuss are: 1) the switch
control signal generation, and 2) possible leakage through non-power pins. A
figure of the \cite{tokunaga2010securing} paper indicates it uses an on-chip
ring oscillator for the switch control clock generation and a programmable
pattern generator for each switching capacitor module. The \cite{gornikhardware}
work exposed the switch controls to chip I/O pins and let a
microprocessor generate the control signals. Supposedly the exposed control
signals are for the testing chip only, for practical defense they should be
implemented on-chip to avoid adversarial manipulations. Since a ring oscillator's
frequency is sensitive to temperature, and so are the clock skews to switches,
it is unclear if switching capacitors would leak information when their
activities are not as well aligned as shown in \reffig{fig:sclclock}. Prior to
\cite{tokunaga2010securing} and \cite{gornikhardware}, Plos
\cite{plos2009evaluation} evaluated early designs using switching capacitors
and identified data-dependent information can also leak through a device's I/O
pins (non-power pin), due to coupling effects (crosstalk) where the switching
activity of one wire influences the voltage of neighboring wires. The
\cite{tokunaga2010securing} paper did not address this problem; attack
measurements were only performed on a serial resistor in the chip's $V_{dd}$
supply line. On the other hand \cite{gornikhardware} cited the Plos paper yet
still did not explicitly test their design's I/O leakage. The power
measurements were performed by a current probe also in the power line only.

\subsection{Noise Generation}

In the time dimension, the most commonly used techniques are randomly inserting
dummy instructions and shuffling the order of operations, and as a result power
traces are not aligned anymore. For example, a microcontroller-based AES cipher
can insert random dummy instructions before and after the S-Box lookup, and the
execution sequence of the 16 S-Boxes is different for
each data block encryption. The price to pay for using dummy instructions is
the throughput of the cipher, and on the other hand shuffling does not affect
the throughput much but the operations that can effectively shuffle are
limited. Multiple other approaches also try to manipulate the clocks of the protected
circuit, including: 1) use clock gating technique to skip randomly clock pulses
\cite{boey2010random}, 2) randomly change the clock frequency
\cite{park2010random} and phase \cite{guneysu2011generic}, 3) use multiple clock
domains for different parts of the protected circuit \cite{bayrak2013eda}.
Another form of temporal distortion coupled with voltage scaling is discussed
in \refsec{sec:dvfs}.

For countermeasures that change the timing behavior of power traces, it is
critical that the attackers can not identify them. Dummy
instructions such as \texttt{NOP}s have detectable low power profile than
others, and the attacker can quickly filter them and restore the alignment of
power traces. Due to the limited degree of shuffling options, the right order
of segments in a power trace can be recovered by signal processing. Pattern
matching techniques are commonly used, such as using a sliding window to
perform correlation tests on one power trace with respect to a reference.
Clavier \etal \cite{clavier2000differential} developed a sliding window based
integration method that can effectively attack shuffled power traces. Newer and
faster alignment techniques have also been published such as the elastic
alignment \cite{van2011improving} and the horizontal alignment
\cite{tian2012general}. In general, the insertion of dummy instructions and
shuffling do not provide a high level of protection against power analysis. A
technique to identify changing clock frequencies is in \refsec{sec:dvfs}.

In the amplitude dimension, the obvious goal is to increase the noise variance
$Var(P_{sw.noise} + P_{el.noise})$ so that the \textit{SNR} drops. Software, if
possible, can run multiple independent threads in parallel. Besides parallelism
hardware architecture also uses dedicated noise engines that perform random
switching activities while the cryptographic circuit operates. A variety of
ways can generate excess noise on the chip. Le Masle \etal \cite{le2011constant}
use long wires with many buffers along the route, and feed each wire with
controlled switching activities to draw currents in the buffers. G{\"u}neysu
\etal \cite{guneysu2010using} find in dual-ported memories write contentions
result in metastability within the storage cells and lead to increased power
consumption. They also developed a way to modify the Xilinx FPGA bitstream to
generate controlled short circuits for a very limited amount of time. It is
important for noise creators to cover uniformly the entire protected circuit
but not only sit in a corner of the chip. Cornered noises may only affect
measurements from nearby power lines, yet still leave backdoors in other
power/ground pins and electromagnetic emissions. Another critical requirement
for amplitude noises is they must have the identical frequency spectrum as the
exploitable power; otherwise they are simply removed by a frequency
filter. Equation \refeq{eq:corsnr} shows the correlation coefficient in CPA is
proportional to $\sqrt{\mathit{SNR}}$, and so empirically added amplitude noise
results in quadratic number of power traces to get the same attack outcomes. In
practice, no countermeasure can make noise variance to infinity, and for CPA
only the relative amplitude of correlation coefficients matters not
the absolute. So amplitude noise is often considered (\cite{mangard2008power}
and \cite{guneysu2011generic}) not an optimum way to thwart power side-channel
attacks. They play more assisting roles in defense.

\subsection{Random Voltage and Frequency Scaling} \label{sec:dvfs}

The Dynamic Voltage and Frequency Scaling (DVFS) technique is generally used for
digital systems to achieve power efficiency. From the defense perspective, it could
be randomized to halt power side-channel attacks. Compared to amplitude noises
that mainly manipulate $C_L$, scrambling both $V_{dd}$ and $f$ is supposedly more
efficient ($P_{dyn} \propto f \cdot C_L \cdot V_{dd}^2$). Yang \etal
\cite{yang2005power} first introduced DVFS for counteracting DPA attacks.
Their proposed cryptosystem consists of three parts: 1) a cryptoprocessor, 2) a
DVFS Feedback Loop (DVFSL), and 3) a DVFS Scheduler (DVFSS). During critical
operations of the cryptoprocessor the DVFSS unit randomly generates ($V_{dd}$,
$f$) configurations and sends them to the DVFSL, which implements the support
for DVFS. The DVFSL unit thereafter can apply the ($V_{dd}$, $f$) changes to
the cryptoprocessor with optional random delays in between. The authors proved
their concept using a software implementation of the DES cipher running by the
SimplePower tool, an RTL power estimator. No actual power analysis attacks were
mounted. Using custom metrics for power trace properties the authors claimed
the effectiveness of randomized DVFS (RDVFS) as a DPA-resistant technique.

\begin{figure}[h]
  \centering
  \includegraphics[width=.45\textwidth]{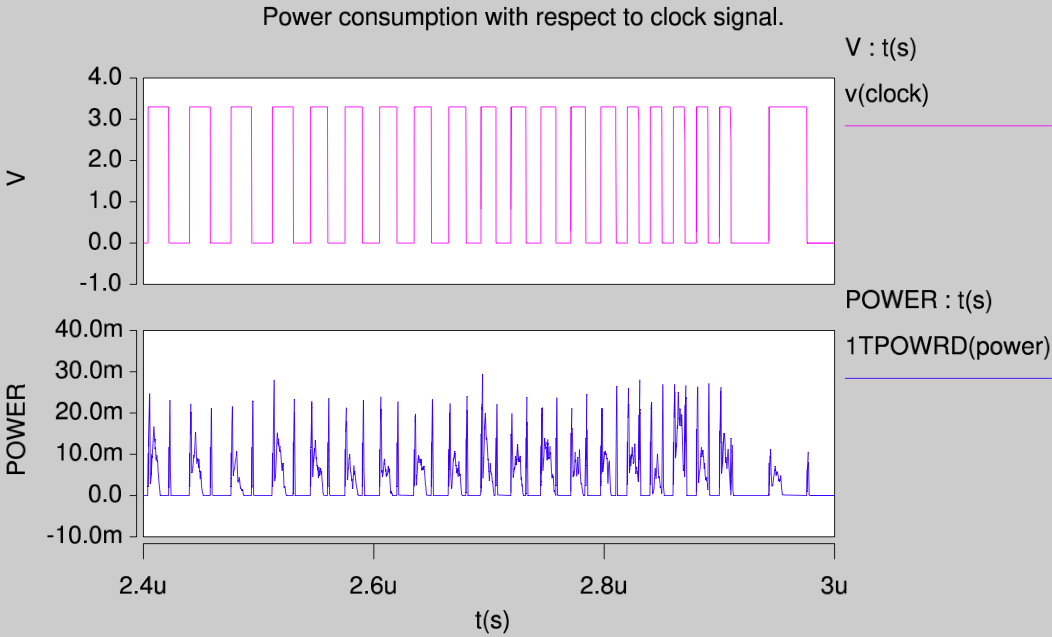}
  \caption{Frequency Detection \cite{baddam2007evaluation} of RDVFS.
           Bottom: Simulated Power Traces with RDVFS.
           Top: The Predicted Clock Frequencies.}
  \label{fig:rdvfs}
\end{figure}

Baddam \etal \cite{baddam2007evaluation} evaluated the concept of
\cite{yang2005power} using more realistic experiment setup and actual DPA
attacks. They implemented an RDVFS-enabled AES (partial) test circuit using an
AMS 0.35-$\mu$m technology library and run the SPICE netlist simulation without
routing parasitics (for simulation speed). The simulated power traces were then
tested using DoM and CPA based attacks. In the power traces, they notice
distinguishable higher spikes, each followed by a slope of setting down towards
the next one (\reffig{fig:rdvfs} Bottom). The reason is most transitions in a
sequential circuit happen right after clock edges and cause a burst of
currents. Therefore, the clock frequencies can be sequenced by reading the
spikes in the power traces (\reffig{fig:rdvfs} Top). Because in DVFS commonly
there is a one-to-one mapping between each operating voltage and frequency, the
voltages of RDVFS can be derived. The attacker then scales power models
accordingly and concludes successful key extraction. This experiment also
confirms that simply manipulating clocks as temporal noises do not provide much
protection. Observing the limitations of RDVFS Baddam \etal tried only to 
scale the voltage while keeping the frequency constant. However for their
experiment with 10,000 encryption rounds the correlation strength was never
below a point where the key was undetectable.

Avirneni and Somani \cite{avirneni2013countering} tried to overcome the
limitations of RDVFS by introducing Aggressive Voltage Scaling (AVS) and
Aggressive Frequency Scaling (AFS). These techniques break the one-to-one
connection between $V_{dd}$ and $f$. For operating frequency beyond the timing
margins, the authors suggest self-correction circuit design. For the evaluation,
they set up SPICE simulation using one S-Box circuit of AES and apply random
($V_{dd}$, $f$) pairs for collecting 10,000 power traces. The traces are
evaluated by CPA based attacks. The same experiment was repeated 500 times
and only two times the actual key had the strongest correlation.

DVFS shows potential for mitigating power side-channel attacks but the results
are still preliminary. For commercial CPUs and FPGAs that already have DVFS
support the random DVFS methods should be feasible options. Otherwise, the
on-chip design of the voltage regulator, clock management, and error-correction
logic is not yet a trivial task for custom ASICs.

\subsection{Combining Countermeasures}

It is widely accepted that a single countermeasure can not effectively
protect the security hardware against a variety of side-channel attacks.
Accordingly, a straightforward thought is to combine protections.
Unfortunately, some countermeasures do not simply add up, as explained by the
folding attack in \refsec{sec:rsl_mdpl}. So choosing a combination of suitable
countermeasures is a rather challenging and tedious work in practice. A
positive example of using combined mitigations is proposed by G{\"u}neysu \etal
\cite{guneysu2011generic} for FPGAs. They use signal masking, clock
randomization, and amplitude noise together for protection. The design could
resist a variety of first-order DPA attacks, with up to 100 million power
traces. Combining masked and precharge/dual-rail logic gates is another popular
direction. Besides the RSL and MDPL, \textit{LUT-Masked Dual-Rail with
Precharge Logic (LMDPL)} \cite{leiserson2014gate} is the most recent tested
hardware implementation by the time of writing this survey. LMDPL uses on
gate-level masking, dual-rail precharge logic without the need for routing
constraints. It employs monotonic gates to avoid glitches, and it fights
against leakage of mask bits and early propagation using a novel
\textit{activity image analysis} for combinational data paths. In its FPGA
evaluation, it shows no significant leaks up to 200 million power traces.

Obviously using multiple mitigations further increases the circuit area, power,
cost and complicates the design flow. Improving the security level is never
free and never perfect. The cost and effort for the defense should match the
value of the device. For a chosen combination of mitigations, it should still
be used with corresponding key updating policies to make allowed power trace
queries of the same key within its tolerance.

\section{Summary}

Making a device resistant to power side-channel attacks is not trivial.  In
academic publications, resistant logic styles have been the most popular form
of hardware countermeasures. Moreover, according to \cite{mangard2008power} in
the semiconductor industry, counteracting power side-channel attacks at the
cell level was one of the first reactions. It is reasonable because logic
styles are close to solving the information leaks fundamentally in power
traces. Hence, this survey reviewed foundational designs using logic style
countermeasure and their vulnerabilities. Since the exposure of flaws in
DPA-resistant logic styles, a large number of proposals have been raised for
improvements, at a higher cost of the area, speed, and power. Some logic styles
such as self-timed (asynchronous) logic design \cite{taylor2002improving} and
current-mode logic \cite{allam2001dynamic} are not discussed because they are
not quite compatible with contemporary mainstream IC design flows.

\begin{table*}[t]
\small
\centering
\begin{tabular}{l|l}
\hline
\textbf{Countermeasures} & \textbf{Potential Vulnerabilities} \\ \hline
\textbf{Masked Logic}         & \begin{tabular}[c]{@{}l@{}}\textbf{Glitches:} Gate toggles before final values settle can leak information.\\ \textbf{Higher-Order Leaks:} Combining samples at multiple times can remove the effect of masks.\\ \textbf{Random Numbers:} Random number generators can be biased to weaken the countermeasure.\end{tabular} \\ \hline
\textbf{DRP Logic}            & \begin{tabular}[c]{@{}l@{}}\textbf{Routing Imbalance:} Imbalanced load and parasitic capacitances cause leakage in power traces.\\ \textbf{Early Propagation:} Gates evaluate output before all inputs settle can leak information.\end{tabular} \\ \hline
\textbf{Masked DRP Logic}     & \begin{tabular}[c]{@{}l@{}}\textbf{Masking Bits:} Masking bits could be detected and removed by a variety of means.\\ \textbf{Routing Imbalance:} Imbalanced load and parasitic capacitances cause leakage in power traces.\\ \textbf{Early Propagation:} Gates evaluate output before all inputs settle can leak information.\\ \textbf{Nonuniform Switchings:} If masking bits are compromised, nonuniform switchings can be attacked.\\ \textbf{Random Numbers:} Random number generators can be biased to weaken the countermeasure.\end{tabular} \\ \hline
\textbf{Switching Capacitors} & \begin{tabular}[c]{@{}l@{}}\textbf{Switches:} Imperfect semiconductor switches can leak information when they are open.\\ \textbf{Control Signals:} Signals controlling capacitor switchings can be biased to weaken the countermeasure.\end{tabular} \\ \hline
\textbf{Temporal Noises}      & \begin{tabular}[c]{@{}l@{}}\textbf{Detectable Patterns:} Signal processing techniques can detect and remove temporal distortions.\\ \textbf{Random Numbers:} Random number generators can be biased to weaken the countermeasure.\end{tabular} \\ \hline
\textbf{Amplitude Noises}     & \begin{tabular}[c]{@{}l@{}}\textbf{Amplitude Variance:} Amplitude variances can not be infinity and have limited effectiveness.\\ \textbf{Frequency Spectrum:} Noises having different frequency components than signals can be filtered.\\ \textbf{Random Numbers:} Random number generators can be biased to weaken the countermeasure.\end{tabular} \\ \hline
\textbf{RDVFS}                & \begin{tabular}[c]{@{}l@{}}\textbf{Detectable Patterns:} Signal processing techniques can detect and remove temporal distortions.\\ \textbf{Frequency-Voltage Bonding:} Once frequencies are detected, voltages can be derived and scaled.\\ \textbf{Random Numbers:} Random number generators can be biased to weaken the countermeasure.\end{tabular} \\ \hline
\end{tabular}
\caption{Existing Hardware Countermeasures and Potential Vulnerabilities}
\label{tab:countermeasures}
\end{table*}

Although numerous logic style countermeasures exist, each of them features a
unique set of overheads and constraints. The lack of security support in CAD
tools has limited the implementation of many defensive circuit techniques.
Hardware defenders have to swap gates, use customized routers and dictate
existing tools for specialized layout requirements. Moreover extra effort on
functional and timing verification has been incurred for secure
implementations. To circumvent complications in the design flow the rest
countermeasures are engaged with separate objects to relax the constraints, for
example, random number generators, clocks, and voltage regulators that
themselves must be tamper-resistant. Once random numbers are biased, or clock
sources are compromised, it is anticipated that corresponding countermeasures
are weakened. \reftab{tab:countermeasures} summarizes the potential
vulnerabilities in existing hardware countermeasures. The report aims to
provide a survey of exposed vulnerabilities in hardware-based countermeasures,
and serve as a reference for future more secure IC implementations.

\section{Acknowledgement}
This work was partially supported by NSF Awards 1018850, and 1228992, and by
C-FAR, part of STARnet, a Semiconductor Research Corporation program. We also
thank the members of Michael Taylor's Bespoke Silicon Group for feedback on the
paper.

\bibliographystyle{alpha}
{\footnotesize
\bibliography{sigproc-sp}}  
%
%

\clearpage
\appendix

\section{The AES Cipher}


The AES \cite{pub2001197} block cipher encrypts 128 bits fixed block size of
data, using a \textit{cipher key} that can be either 128, 192 or 256 bits long.
The differences between the uses of the three key sizes are the number of
encryption/decryption \textit{rounds} and the \textit{key expansion} process.
The three different AES key sizes use 10, 12, and 14 rounds respectively. This
paper mainly discusses the AES-128 version that uses a 128-bit cipher key.

\subsection{The Round Functions}
\begin{figure}[h]
  \centering
  \includegraphics[width=.45\textwidth]{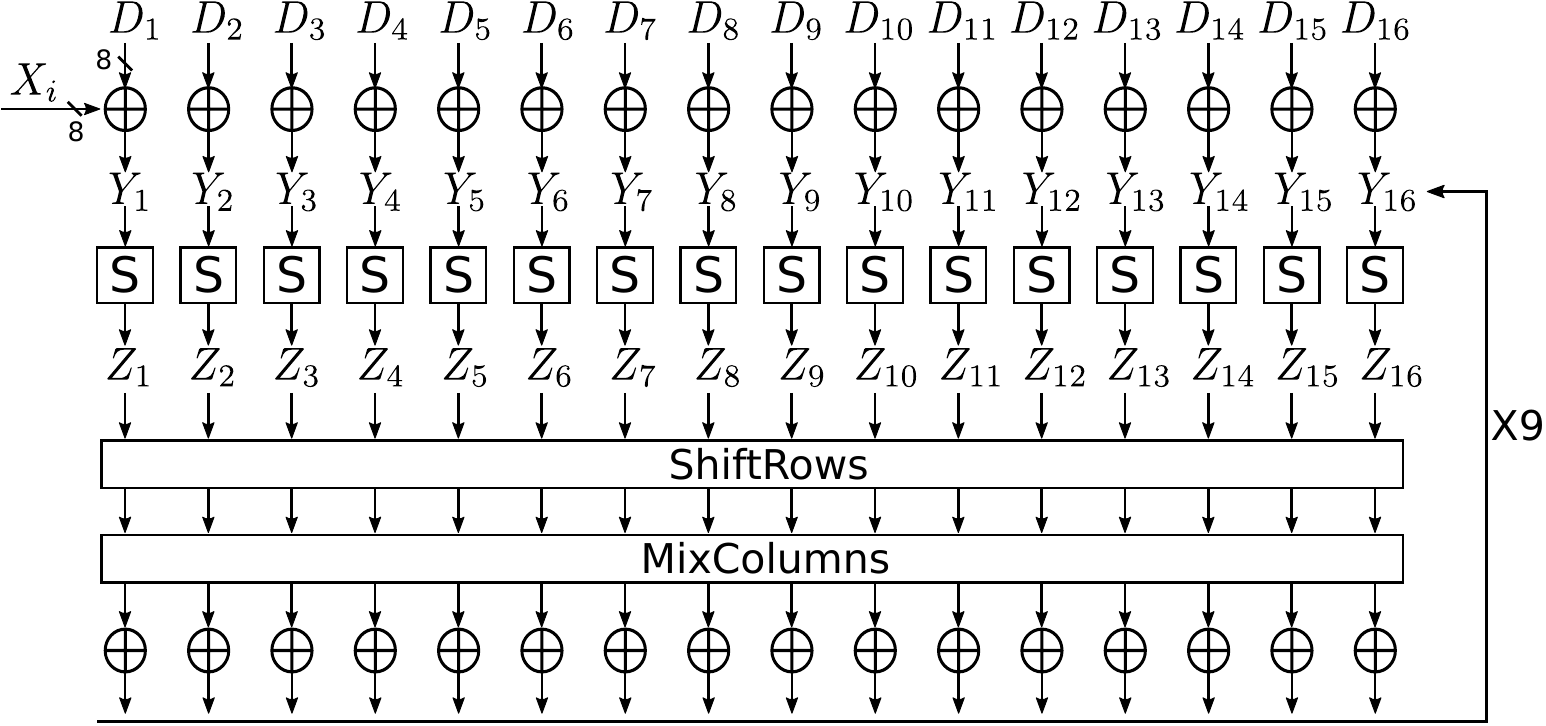}
  \caption{The AES Encryption}
  \label{fig:aes}
\end{figure}

Each round of the AES encryption comprises of four different functions, named
as \textit{AddRoundKey}, \textit{SubBytes}, \textit{ShiftRows}, and
\textit{MixColumns}. The 128-bit data being operated in each round contains 16
bytes \{$D_1$, $D_2$, $\cdots$, $D_{16}$\}, known as the \textit{state}. An
AES state is stored as intermediate values in memory units for subsequent
transformations. The encryption process of one data block starts with the
plaintext state and performs an initial AddRoundKey operation. Then nine rounds
are executed, and each round sequentially performs SubBytes, ShiftRows,
MixColums, and AddRoundKey on the state. After that, a \textit{last round} is
executed with only three functions: SubBytes, ShiftRows, and AddRoundKey. The
encryption flow of AES is in \reffig{fig:aes}, and each function of an AES
round is briefly explained below.

\textbf{AddRoundKey:} In AES-128 a \textit{round key} is also a 128-bit value
that is derived from the cipher key by the key expansion algorithm. Round keys
are different in each round. The AddRoundKey function performs a bitwise XOR
($D_i \oplus X_i$, $i \in [1 \cdots 16]$) of a state byte $D_i$ and a
corresponding round key byte $X_i$. Note that the round key used in the initial
AddRoundKey operation is identical to the cipher key.

\begin{equation} \label{eq:subbytes}
  S(Y_i) = \mathbf{A} \cdot Y_i^{-1} + \mathbf{b}
\end{equation}

\textbf{SubBytes:} Also known as the S-Box, this function is to perform byte
substitution on each byte of the state \textit{independently} after the
AddRoundKey operation. The function \refeq{eq:subbytes} for each state byte
($Y_i$) is to compute its multiplicative inverse ($Y_i^{-1}$) in the finite field
of $GF(2^8)$, followed by an affine transform. Definitions of binary matrix
$\mathbf{A}$ and vector $\mathbf{b}$ are given by \cite{pub2001197}.
Due to the computational complexity of finite field inversions and matrix
multiplications, the S-Box is often precomputed as a lookup table and stored in
memories for quick reference in each round.

\textbf{ShiftRows:} This transformation views the 16 state bytes as a 4
$\times$ 4 array and performs cyclic left shifts for each row of the state
bytes. The first state row is not shifted, the second row is shifted to the
left by one byte position, the third row is shifted by two byte positions and
the fourth row is shifted by three.

\textbf{MixColumns:} This column-wise function provides further mixing of bytes
in a state. It produces a new state column using a constant matrix defined by
AES to multiply a previous state column.


The state gets updated by each round function, and after ten rounds of
transformations the initial 128-bit plaintext is turned into the 128-bit
ciphertext. More plaintext bits are encrypted in the same way as 128-bit data
blocks. The AES decryption process executes the inverse of these functions in
reversed order, using the \textit{same cipher key}. The complete specification
of the AES is in the FIPS-197 document \cite{pub2001197}. The
provided information in this section should suffice most discussions covered in
the survey.


The SubBytes function has the property that small changes in its input produce
large differences in the output, and vise versa. Therefore, most power analysis
attacks target the S-Box input or output because wrong key guesses
lead to clearly distinguishable S-Box outputs compared to the actual values. In
contrast, cryptographically weak functions such as the AddRoundKey are more
resistant to power analysis because similar key guesses result in small
variations in these functions.

\subsection{The Cipher Implementations}


The AES cipher can be efficiently implemented in either software or hardware.
Many smart cards have software implementations using 8051, PIC or ARM family
microcontrollers \cite{rankl2010smart}. A simple microprocessor with 8-bit
data paths does not create any significant limitations for running the AES
algorithm. Some modern smart cards could also use 32-bit microprocessors for
advanced features such as running Java programs. For ultra low-power,
performance, and some security advantages, hardware implementations of cipher
modules are also found available. Instead of stored lookup tables, the S-Box
can also be efficiency implemented in hardware using finite field arithmetic
\cite{wolkerstorfer2002asic}. There is usually more parallelism in the
hardware implementation than in software. For example, up to 16 S-Boxes can be
executed simultaneously, while an 8-bit microcontroller can only process one
byte after another. Both software and hardware ciphers often lack
countermeasures for power analysis attacks unless specially designed for such
purposes.


The cipher was designed to resist algorithmic breaking attempts: brute-forcing
$2^{128}$ guesses is computationally prohibitive. However despite its software
or hardware implementations, it is obvious that its internal computations often
operate on smaller data widths than 128 bits, especially the S-Box function
that only works on 8-bit chunks independently. It should be emphasized that
these intermediate values are exactly on target in power analysis attacks, and
their short length in bits greatly reduces the effort of exhaustive testing.

\balancecolumns
\end{document}